\documentclass[11pt]{article}
\usepackage{jheppub}
\usepackage{epsfig}
\usepackage[utf8]{inputenc}
\usepackage{graphicx,tikz}
\usepackage{amsfonts,amsmath,amssymb,mathrsfs}
\usepackage{array,mathrsfs,yfonts,dsfont,bbm,colonequals,amscd,mathtools}
\usepackage{relsize,suffix,cancel,bbm,capt-of,upgreek,verbatim}
\usepackage{cleveref}
\usepackage{comment}

\usepackage{hyperref}
\usepackage{multirow}
\usepackage[font=small,labelfont=bf,margin=0.05\textwidth]{caption}
\usepackage{mdframed}
\usepackage[mode=buildnew]{standalone}

\usepackage{diagbox} 

\newcommand{\beq}{\begin{equation}}
\newcommand{\eeq}{\end{equation}}
\newcommand{\bba}{\begin{align}}
\newcommand{\eea}{\end{align}}
\newcommand{\beqq}{\begin{equation*}}
\newcommand{\eeqq}{\end{equation*}}
\newcommand{\Res}{\operatorname*{Res}}

\newcommand{\OOk}{\mathcal{O}}
\newcommand{\OOxy}{\mathsf{O}}

\title{Ten dimensional symmetry of $\mathcal{N}=4$ SYM correlators}

\author{Simon Caron-Huot,}
\emailAdd{schuot@physics.mcgill.ca}
\author{Frank Coronado}
\emailAdd{fcidrogo@gmail.com}

\affiliation{Department of Physics, McGill University, 3600 Rue University, Montr\'eal, QC Canada H3A 2T8}

\abstract{
We consider four-point correlation functions of protected single-trace scalar operators
in planar $\mathcal{N}=4$ supersymmetric Yang-Mills (SYM).
We conjecture that all loop corrections derive from an integrand
which enjoys a ten-dimensional symmetry.
This symmetry combines spacetime and R-charge transformations.
By considering a 10D light-like limit, we extend the correlator/amplitude duality by equating
large R-charge octagons with Coulomb branch scattering amplitudes.
Using results from integrability, this predicts new finite amplitudes as well as some
Feynman integrals.
}

\begin{document}
\maketitle
\flushbottom

\section{Introduction}

The loop integrands of (some) correlation functions in planar $\mathcal{N}=4$ SYM have hidden symmetries that highly constrain their form.
The most interesting and studied example is the four-point correlator of scalars in the stress tensor multiplet.
Because loop corrections can be obtained by inserting the Lagrangian density, which is part of the same multiplet, the loop integrand enjoys a full permutation symmetry which treats the four external points and loop integration points on the same footing.
This powerful symmetry has played a key role in ``bootstrapping'' the integrand \cite{Eden:2011we,Eden:2012tu}
(now up to ten loops \cite{Bourjaily:2016evz}) by imposing various limits on an Ansatz of conformal integrals.

The stress tensor multiplet is the first of an infinite tower of protected (half-BPS) single-trace operators present in this theory. These 
have scaling dimension $\Delta=k$ 
with $k\geq 2$ and can be interpreted, in the 
AdS${}_5\times$S$_5$ dual realization of this theory, as Kaluza-Klein modes of the graviton on the five-sphere.
They thus offer a unique window into the 10-dimensional nature of the dual theory.
The combinatorics of their correlators $\mathcal{G}_{k_1 k_2 k_3 k_4}$
however becomes rapidly complicated because of the proliferation of R-charge structures.
Furthermore, the full permutation symmetry that interchanges external and internal points is lost: the bootstrap problem has many more free parameters to fix.  Nevertheless, with the discovery of a recursive relation between different integrands in their light-cone limit, all integrands were obtained up to three loops \cite{Chicherin:2015edu} and later to five loops using further input from integrability \cite{Chicherin:2018avq}. These results showed that each integrand $\mathcal{G}_{k_1 k_2 k_3 k_4}$ contains only a subgroup of the basis of conformal integrals from the stress tensor case $\mathcal{G}_{2222}$ and, not surprisingly, this subset gets smaller as the charges $k_{i}$ increase. 

In this paper we propose that the integrands $\mathcal{G}_{k_1 k_2 k_3 k_4}$
all follow from a single generating function which enjoys a ten-dimensional hidden symmetry.  At ``low'' loop orders (up to at least seven loops),
this generating function can be uniquely uplifted from the known stress-tensor case
by replacing all four-dimensional distances $x_{ij}^{2}$
by ten-dimensional ones $X_{ij}^{2} \equiv x_{ij}^2+y_{ij}^{2}\,$,
where the $y_i$ are null 6-vectors parametrizing the R-charge of the external operators.
By expanding at small $y_{ij}^2$ and selecting terms with the correct charge, we have verified that this generating function reproduces precisely all
$\mathcal{G}_{k_1 k_2 k_3 k_4}$ in the literature.

As an application, we observe that the ten-dimensional symmetry equates two natural limits in which correlators simplify or factorize: when an edge becomes ``fat'' from carrying a large R-charge, which prevents propagators from crossing it (at finite loop order), or when an edge becomes null, creating a high-energy excitation that can't be deflected.  Both are realized simply by taking the 10-dimensional null limit $X_{ij}^2\to 0$.
Combining this observation with the known relation between null limits and scattering amplitudes, we conjecture a new relation between scattering amplitudes on the Coulomb branch and the 10D null limit of charged correlators.

This conjecture implies that the so-called \textit{octagon},
a family of four-point functions with large R-charge \cite{Coronado:2018ypq,Coronado:2018cxj},
has the same integrand as the four-gluon amplitude.
When restricted to a suitable subset of the Coulomb branch,
on which infrared divergences of the amplitudes cancel, the integrated expressions also coincide.
In other words, the octagon is a scattering amplitude.
Exact results from integrability \cite{Kostov:2019stn,Kostov:2019auq}
 then allow to predict new integrated expressions for various (combinations of) Feynman integrals.

This paper is organized as follows. In section~\ref{sec:ReviewInt} we review known results on the loop integrands of half-BPS correlators, defined through the Lagrangian insertion method. In section~\ref{sec:TenSym} we propose generating functions which unite them, first for free correlators and then for loop integrands. We make explicit the ten-dimensional symmetry of these generating functions and show that they uniquely uplift from the stress tensor case up to at least seven loops.
In section~\ref{sec:NewDuality} we consider a ten-dimensional null limit which allows us to establish a novel duality between correlators with large R-charge (octagons) and massive scattering amplitudes on the Coulomb branch. We argue this duality holds for their integrands and also at the integrated function level.  We observe that different ways to approach the four-dimensional massless limit are controlled by exponents which differ from the cusp anomalous dimension. In section~\ref{sec:Integrability} we use recent integrability results to provide a finite coupling representation for the octagon/amplitude. We focus on the weak coupling limit where we can compare with scattering  amplitude integrals. This allows us to rediscover the map between regular fishnet diagrams and determinants of ladder integrals, as well as extending it to include deformations of the fishnets. In section~\ref{sec:discussion} we discuss our results and possible extensions. In appendix~\ref{app:3loops} we record a basis of three-loop planar integrals used in the main text. Finally, in appendix~\ref{app:5loopIdentity}, we exemplify a five-loop identity derived as a consequence of the duality.


\section{A review of correlators and loop integrands}
\label{sec:ReviewInt}

We consider $\mathcal{N}=4$ super Yang-Mills with gauge group SU($N_c$) (or U($N_c$)).
In its gauge theory formulation, the theory contains six scalar fields $\Phi=(\phi^{1},\cdots, \phi^{6})$
which are all $N_c\times N_c$ Hermitian matrices (traceless in the SU($N_c$) case).
We will be interested in single-trace half-BPS protected operators, 
which are symmetric traceless (with respect to SO(6)) products of these scalars:
\beq\label{eq:Ok}
\OOk_{k_{i}}(x_{i})= \frac{1}{k_{i}}\,\left(\frac{2}{4\pi^2\,N_c}\right)^{k_{i}/2}\text{Tr}\big[(y_{i}\cdot \Phi(x_{i}))^{k_{i}}\big]
\eeq
where we introduced null 6-vectors $y_{i}\cdot y_{i}=0$ in order to
automatically project out the traces.
The chosen normalization, including $1/k$ which accounts for the cyclic symmetry of the trace in $\OOk_{k}$,
will simplify future formulas.
The operator $\OOk_{k_{i}}(x_{i})$ is the lowest-dimensional component of a supermultiplet
which also contains fermions, which will not be considered in this paper.
To fix our notations, the two-point function in the planar limit is
\beq
\langle \OOk_{k}(x_1)\OOk_{k}(x_2)\rangle = 
\left(\frac{1}{k}+O(1/N_c^2)\right)(d_{12})^k \quad\mbox{where}\quad d_{ij}\equiv \frac{2\, y_{i}.y_{j}}{x_{ij}^2}= \frac{-y_{ij}^2}{x_{ij}^2}
\eeq
with $x_{ij}=x_i-x_j$.\footnote{In the literature, $y_{ij}^2$ is sometimes defined
as a dot product: $(y_{ij}^{2})^{\rm elsewhere}=y_i{\cdot}y_j$.  Here it will be advantageous to
treat $x$ (spacetime) and $y$ (R-charge) variables symmetrically: $(y_{ij}^2)^{\rm here}\equiv (y_i-y_j)^2$.
}

\subsection{Correlators and integrands}

The perturbative series of the connected planar four-point function takes the generic form:
\beq
N_c^2\langle  \OOk_{k_{1}}\OOk_{k_{2}} \OOk_{k_{3}} \OOk_{k_{4}} \rangle_{\rm c} \, =
\,
G^{\rm free}_{k_1 k_2 k_3 k_4} + \sum_{\ell=1}^{\infty} G^{(\ell)}_{k_1 k_2 k_3 k_4}  +O(1/N_c^2)\,,
\eeq
where the loop corrections can be computed using the Lagrangian insertion method \cite{Intriligator:1998ig,Eden:2011we}:
\beq\label{eq:Gcorrelator}
G^{(\ell)}_{k_{1}k_{2}k_{3}k_{4}}
= \frac{(-g^2)^\ell}{\ell!}\int \frac{d^4x_{5}}{\pi^2} \cdots \frac{d^4x_{4+n}}{\pi^2} \, \mathcal{G}^{(\ell)}_{k_{1}k_{2}k_{3}k_{4}}\,,
\eeq
with $g^2\equiv \frac{g^2_{\rm YM}N_c}{16\pi^2}$ the coupling constant.
All our integrals are written after Wick rotation to Euclidean signature.
The $\ell$-loop integrand $\mathcal{G}$ is given by a $(4+\ell)$-point correlator of the four external operators and $\ell$ Lagrangian densities evaluated at Born level
\beq\label{eq:GintegrandLagrangian}
\mathcal{G}^{(\ell)}_{k_{1}k_{2}k_{3}k_{4}} = \langle \OOk_{k_{1}}\OOk_{k_{2}} \OOk_{k_{3}} \OOk_{k_{4}}\mathcal{L}(x_{5})\cdots \mathcal{L}(x_{4+\ell})\rangle^{(0)}\,.
\eeq
The superscript indicates that the correlator itself is computed only in the Born-level approximation which starts at order $g^{0}$.

A partial non-renormalisation theorem imposes a factorization for the four-point correlator and its integrand \cite{Eden:2000bk}:
\beq\label{eq:susyward}
\mathcal{G}^{(\ell)}_{k_{1}k_{2}k_{3}k_{4}} = R_{1234}\left(2\,x_{12}^2x_{13}^2x_{14}^2x_{23}^2x_{24}^2x_{34}^2\right) \mathcal{H}^{(\ell)}_{k_{1} k_{2} k_{3} k_{4}} \,.
\eeq
The first factor has conformal weight $(-1)$  and harmonic weight $2$ at each point:
\bba \label{R}
R_{1234}&= d_{13}^2 d_{24}^2 x_{13}^2x_{24}^2 + d_{12}d_{23}d_{34}d_{14}\left(x_{13}^2x_{24}^2-x_{12}^2x_{34}^2-x_{14}^2 x_{23}^2\right) +(1\leftrightarrow 2) \, + \, (1\leftrightarrow 4)\,.
\end{align}
The reduced integrand $\mathcal{H}$ then transforms like a correlator of operators of charge $(k_i-2)$ and dimensions $(k_i+2)$. In particular, the reduced stress-tensor correlator
$\mathcal{H}_{2222}$ is independent of $y$ variables.

In the literature, a generic $\mathcal{H}$ is typically expanded as a sum of
various R-charge structures distinguished by the exponents of $d_{ij}$: 
\beq\label{eq:fromHtoF}
\mathcal{H}^{(\ell)}_{k_{1}k_{2}k_{3}k_{4}} =  \sum_{\{b_{ij}\} \atop k_{i} =2+\sum_{j} b_{ij}}\,  \mathcal{F}^{(\ell)}_{\{b_{ij}\}}
\times \prod_{1\leq i<j \leq 4} \left(d_{ij}\right)^{b_{ij}}
\eeq
where the six-tuples $\{b_{ij}\}\equiv \{b_{12},b_{13},b_{14},b_{23},b_{24},b_{34}\}$ satisfy the constrains  $k_{i} =2+\sum\limits_{j\neq i} b_{ij}$.
This can be interpreted as listing every possible
tree-level R-symmetry structure (products of $d_{ij}$'s) and dressing them with loop corrections.
Importantly, the integrands $\mathcal{F}_{\{b_{ij}\}}$ are rational functions. They
depend exclusively on $x_{ij}^2$ and have the following three defining properties:
\begin{itemize}
\item $\mathcal{F}$ has conformal weight 4 at each point.
\item $\mathcal{F}^{(\ell)}$ has an $S_{\ell}$ symmetry on the integration points $x_{5},\cdots,x_{\ell+4}$.
\item As a consequence of a OPE analysis, $\mathcal{F}$
contains at most a single pole in each $x_{ij}^2$.
\end{itemize}

These properties allow one to make an ansatz for $\mathcal{F}$ with a finite number of parameters.  
At one loop, for example, these conditions are restrictive enough to fix all these structures to be given by the same simple product:
\beq
\mathcal{F}^{(1)}_{\{b_{ij}\}} = \frac{1}{\prod\limits_{1\leq i < j\leq 5}x_{ij}^2}\qquad\text{for all }b_{ij}\geq 0\,. \label{one loop}
\eeq
The fact that the proportionality constant is the same for all R-charges follows from simple factorization limits which will be reviewed below.
At higher loops, more conditions, coming for example from Euclidean or Lorentzian OPEs,
are required to uniquely bootstrap the integrands. In what follows we review the results on the integrand of stress tensor multiplets, which has been pushed up to ten loops, and we review analogous results for charged correlators.

\subsection{The stress tensor multiplet}

The case $k_{i}=2$ is special since as noted the reduced integrand $\mathcal{H}_{2222}$ is independent of $y_{ij}^2$ and thus consists of a single structure at each loop order (see eq.~\eqref{eq:fromHtoF}):
\beq
\mathcal{H}^{(\ell)}_{2222} = \mathcal{F}^{(\ell)}_{\{0,0,0,0,0,0\}}\,.
\eeq
In addition, apart from the obvious permutation symmetries $S_{4}\times S_n$ which permute the external and internal points separately, this object enjoys a remarkable full permutation symmetry $S_{4+n}$ which interchanges external and internal points \cite{Eden:2011we}. This special symmetry stems from the fact that
$\OOk_{2}$ and the Lagrangian density $\mathcal{L}$ are parts of the same supermultiplet, being respectively its bottom and top ($\sim \theta^4$) components.  

Up to three loops, the extended permutation symmetry and planarity are strong enough to fix the integrands up to an overall constant \cite{Eden:2012tu}:
\begin{align}
\mathcal{H}^{(1)}_{2222} \,&=\, \frac{1}{\prod_{1\leq i < j \leq 5}x_{ij}^2}\,, \label{eq:H2oneloop}  \\
\mathcal{H}^{(2)}_{2222} \,&=\,\frac{1}{48} \frac{x_{12}^2x_{34}^2x_{56}^2 \,+\, S_{6}\,\text{permutations}}{\prod_{1\leq i < j \leq 6}x_{ij}^2}\,, 
  \label{eq:H2twoloop}  \\
\mathcal{H}^{(3)}_{2222} \,&=\, \frac{1}{20}\frac{(x_{12}^2)^2\left(x_{34}^2x_{45}^2 x_{56}^2 x_{67}^2 x_{73}^2\right) \,+\, S_{7}\,\text{permutations}}{\prod_{1\leq i < j \leq 7}x_{ij}^2} \,. \label{eq:H2threeloop} 
\end{align}
Notice that each of these is seeded by a single monomial by summing
over $S_{4+n}$ permutations of all points $x_{1},\cdots x_{4+n}$.
The shown denominators simply remove equivalent permutations, e.g. the two-loop integrand in \eqref{eq:H2twoloop} consists of 15 distinct terms with unit coefficient.

These integrands determine, through eqs.~\eqref{eq:Gcorrelator} and \eqref{eq:susyward},
the two-loop correlator:
\bba\label{eq:G2loops}
G^{(1)}_{2222} &= -2\,g^2 R_{1234} \times g_{1234}\,, \\
G^{(2)}_{2222} &= 2\,g^4R_{1234}\,\bigg(h_{12;34}+h_{34;12}+h_{14;23}+h_{23;14}+h_{13;24}+h_{24;13}\nonumber\\
 &\qquad\qquad\; \left. \qquad+ \frac{1}{2} \left(x_{12}^2x_{34}^2 + x_{13}^2 x_{24}^2 + x_{14}^2 x_{23}^2\right)\,[g_{1234}]^2\right)\,,
\end{align}
in terms of one-loop and two-loop ladder integrals:
\bba\label{eq:1-2loopsCI}
g_{1234} &= \frac{1}{\pi^2} \, \int \,\frac{d^{4}x_{5}}{x_{15}^2 x_{25}^2 x_{35}^2 x_{45}^2} \, =\, \frac{F_{1}(z,\bar{z})}{x_{13}^2 x_{24}^2} \,,\nonumber\\
h_{13;24} & = \frac{x_{24}^2}{\pi^4} \int \frac{d^{4}x_{5}d^{4}x_{6}}{(x_{15}^2x_{25}^2x_{45}^2)x_{56}^2(x_{26}^2x_{36}^2x_{46}^2)}\,=\, \frac{F_{2}(z,\bar{z})}{x_{13}^2 x_{24}^2}\,.
\end{align}
The function $F$ is recorded in eq.~\eqref{eq:Fladder} below and the cross-ratios satisfy:
\beq\label{eq:crossratios}
z\bar{z}=\frac{x_{12}^2x_{34}^2}{x_{13}^2x_{34}^2}\,,\quad\,
(1{-}z)(1{-}\bar{z})=\frac{x_{23}^2x_{14}^2}{x_{13}^2x_{34}^2}\,.
\eeq
Similarly, R-charge cross ratios $\alpha,\bar{\alpha}$ are defined using $y_{ij}^2$.

At four and higher loops, the basis of homogeneous polynomials grows as: $3,7,36,220,\cdots$. In order to fix the coefficients of these polynomials in an Ansatz, one needs more conditions on the integrand or integrated correlator. This was achieved by the analysis of kinematical limits which are constrained by factorization,
what is called the soft-collinear bootstrap \cite{Bourjaily:2011hi,Bourjaily:2015bpz}.
As a simple illustration, if $\{ x_{12}^2,x_{23}^2,x_{34}^2,x_{41}^2\}\to 0$ describes a null square, then in the limit that $x_5$ approaches an edge, the most singular term factorizes:  $\mathcal{H}^{(\ell)}\to (\frac{2x_{13}^2x_{24}^2}{x_{15}^2x_{25}^2x_{35}^2x_{45}^2})\times \mathcal{H}^{(\ell-1)}$.
Physically, this ensures the exponentiation of collinear divergences of a dual scattering amplitude.
These lead to linear equations that must be satisfied by the coefficients in the Ansatz.
Such relations and generalizations allowed to
completely fix the planar integrand now up to 10 loops \cite{Eden:2012tu,Bourjaily:2015bpz,Bourjaily:2016evz}.

\subsection{Higher BPS operators and saturation}

\begin{figure}[t]
 \begin{minipage}[h]{1\textwidth}
 \centering 
  \resizebox{.8\totalheight}{!}{\includestandalone[width=1\textwidth]{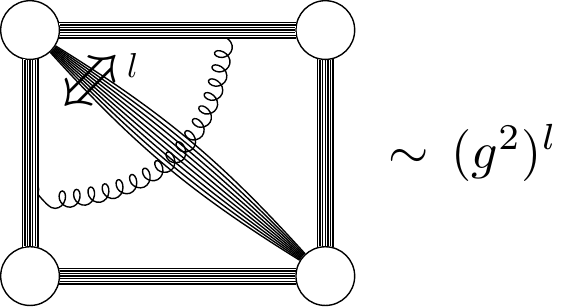}}
 \end{minipage}
 \caption{Saturation: an R-charge factor $(d_{ij})^l$ requires a ``bridge'' made of $l$ propagators and
 cannot be crossed without paying a factor $(g^2)^l$.}
 \label{fig:fatbridge}
 \end{figure}

 \begin{figure}[t]
 \begin{minipage}[h]{1\textwidth}
 \centering 
  \resizebox{1\totalheight}{!}{\includestandalone[width=1\textwidth]{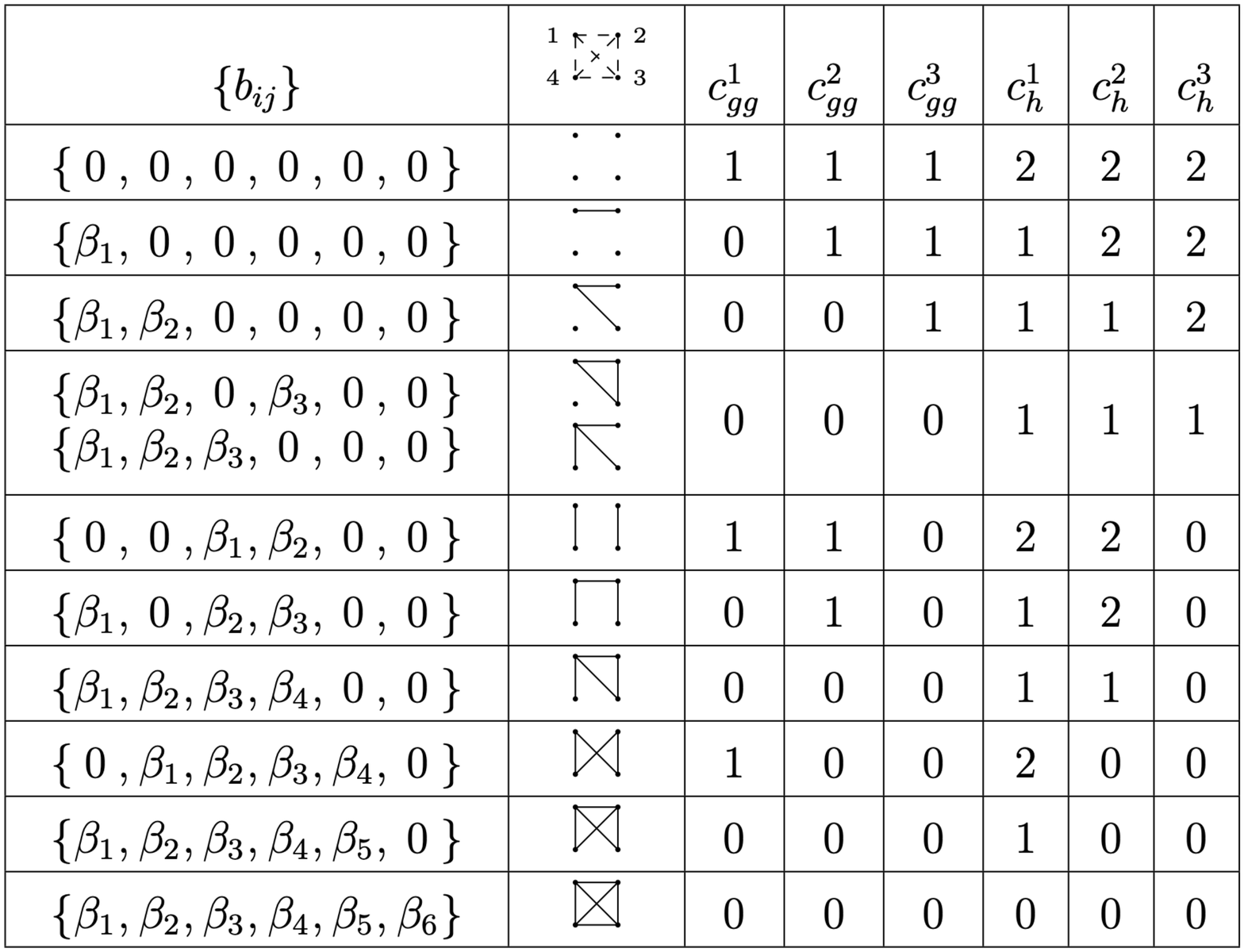}}
 \end{minipage}
 \caption{The full set of two-loop integrands $\mathcal{F}_{b_{ij}}$ according to \cite{Chicherin:2015edu}, 
showing only inequivalent $R$-charge structures.
The coefficients enter eq.~\eqref{twoloop_cs}.
All values $\beta_{i}\geq 1$ are equivalent due to the saturation phenomenon.
 }
 \label{tab:2loops}
 \end{figure}

Moving to charged correlators, the number of structures $\mathcal{F}_{\{b_{ij}\}}$ to determine would appear to be infinite, however the number of distinct functions at a given loop order is finite.
This is due to a saturation bound: if a propagator becomes too ``fat'' it cannot be crossed, as depicted below in fig.~\ref{fig:fatbridge}. Adding more propagators will not change the result, leading to
\beq
\mathcal{F}^{(\ell)}_{\{b_{12},\cdots\}}\equiv \mathcal{F}^{(\ell)}_{\{\kappa_{n},\cdots\}} \quad\mbox{if } b_{12}\geq \kappa_{n}\,.
\eeq
Up to five loops saturation happens at $\kappa_{n}=n-1$, according to \cite{Chicherin:2015edu,Chicherin:2018avq}, however this can change at higher loop orders. Another useful fact is that singularities from propagator poles are also ``uncrossable'',
meaning that residues saturate already at $b_{12}=0$:
\beq
{\rm Res}_{x_{12}^2=0} \mathcal{F}_{\{b_{ij}\}} \,=\, {\rm Res}_{x_{12}^2=0}\,\mathcal{F}_{\{b_{ij}\}}\big{|}_{b_{12}\to 0}\,.
\eeq
These two conditions will be important below.
As discussed in ref.~\cite{Chicherin:2015edu}, they explain much of the structure of R-charged integrands.
As an example we display here the two-loop results for the integrated structures $\mathcal{F}_{\{b_{ij}\}}$ from that reference, in terms of the integrals entering in \eqref{eq:G2loops} and with coefficients recorded in fig.~\ref{tab:2loops}:
\bba
x_{12}^2x_{13}^2 x_{14}^2x_{23}^2 & x_{24}^2x_{34}^2\,\int \frac{dx_{5}^4\,dx_{6}^4}{2!(\pi^2)^2}\,\mathcal{F}^{(2)}_{\{b_{ij}\}} =\nonumber\\
&c^1_h\,h_{12;34} +c^2_h\,h_{13;24}+c^3_h\,h_{14;23}
+ \frac{1}{2}\big{(}c^{1}_{gg}\,x_{12}^2 x_{34}^2+c^{2}_{gg}\,x_{13}^2 x_{24}^2+c^{3}_{gg}\,x_{14}^2 x_{23}^2\big{)} \left[g_{1234}\right]^2 \label{twoloop_cs}\,.
\end{align}
In the next section, we explain how to fully recover the information in such tables
from a concise ten-dimensional integrand.



\section{Ten-dimensional hidden symmetry}
\label{sec:TenSym}

Our goal is now to understand how different R-charge correlators fit together as a single object.
The key idea will be to define a ``master'' operator given as a sum of all scalar BPS operators:
\beq \label{def O}
\OOxy(x,y) \,\equiv\, \sum_{k=1}^{\infty}\,\OOk_{k}(x,y)\,.
\eeq
The operator $\OOk_{1}$ vanishes when we consider $SU(N_{c})$ gauge group, in which case
the sum starts effectively from $\OOk_{2}$.  For the free correlators it will be more convenient
to work in the $U(N_c)$; formulas for $SU(N_c)$ are then trivially obtained by subtracting off terms linear in $y$.  For the loop integrands, there will be no differences between $U(N_c)$ and $SU(N_c)$.

To create a generating function of R-charged correlators, we simply evaluate correlation functions
of these master operators:
\beq
\langle \OOxy(x_1,y_1)\,\OOxy(x_2,y_2)\cdots \rangle \,.
\eeq
As a warm-up, we first perform this task for the free theory, starting with three- and four-point functions.
After stating our main conjecture in subsection \ref{ssec:conjecture},
we discuss loop integrands as well as integrated correlators in \ref{sec:Integrated23}.

\subsection{All free three-point and four-point functions}

The three-point functions of our BPS operators are protected so the free theory result is exact and it is given by a single R-charge structure \cite{Lee:1998bxa}:
\beq \label{3pt}
\langle\OOk_{k_{1}}\,\OOk_{k_{2}}\,\OOk_{k_{3}}\rangle\,=\, \frac{1}{N_c}\,\left(d_{12}\right)^{l_{12}}\,\left(d_{23}\right)^{l_{23}}\,\left(d_{31}\right)^{l_{31}}
\eeq
with $l_{12} = \frac{k_{1}+k_{2}-k_{3}}{2}\geq 0$ and cyclic rotations of this for $l_{23}$ and $l_{31}$. Our normalization in \eqref{eq:Ok}, with $1/k$
 for each $\mathcal{O}_{k}$, conveniently cancels the cyclic permutations of each operator. This was chosen so that the three-point functions will sum up to a geometric series.

\begin{figure}[t]
 \begin{minipage}[h]{1\textwidth}
 \centering 
  \resizebox{1.75\totalheight}{!}{\includestandalone[width=1\textwidth]{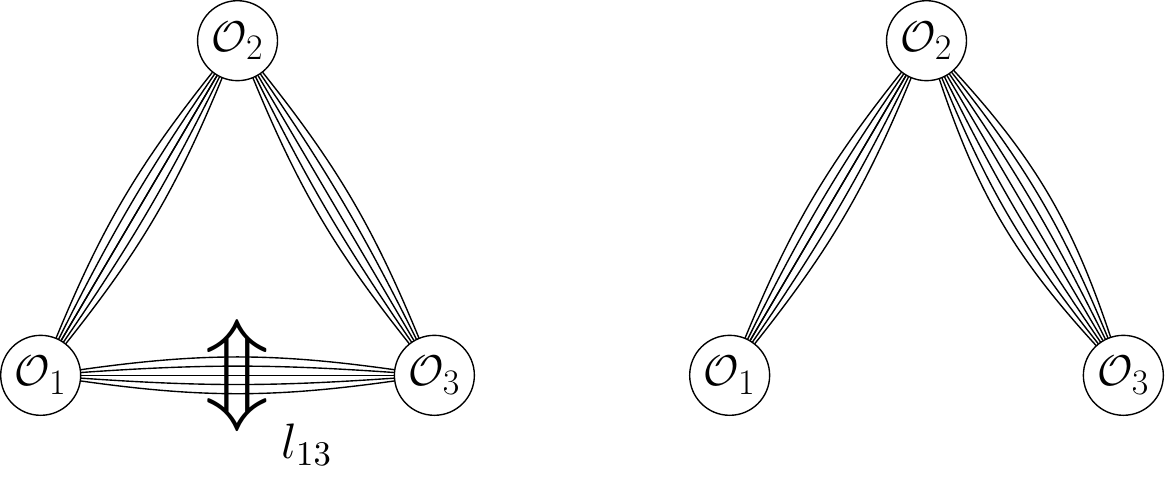}}
 \end{minipage}
 \caption{Generic versus extremal three-point functions.  The second type can be removed
 by redefining $\mathcal{O}$'s to be orthogonal to double-trace operators.}
 \label{fig:ThreePoint}
\end{figure}

The generating function of all three-point functions is computed as
\beq
\langle \OOxy(x_1,y_1)\,\OOxy(x_2,y_2)\,\OOxy(x_3,y_3) \rangle\, = \sum_{l_{ij}\geq 1} 
\langle\OOk_{k_{1}}\,\OOk_{k_{2}}\,\OOk_{k_{3}}\rangle
+ \frac{1}{N_c}\mathbb{T}_{\text{extr}} \label{eq:GG3point}
\eeq
where we separated generic terms from those with zero-length bridge, 
often called ``extremal'' correlators, see fig.~\ref{fig:ThreePoint}.
Introducing a new effective propagator:
\beq
D_{ij} \equiv \frac{d_{ij}}{1-d_{ij}} =\frac{-y_{ij}^2}{X_{ij}^2}
\eeq
which involves the ten-dimensional distance that combines spacetime and R-charge space,
\beq
X_{ij}^2 \equiv x_{ij}^2 + y_{ij}^2\,,
\eeq
the correlator becomes simply
\beq
\langle \OOxy(x_1,y_1)\,\OOxy(x_2,y_2)\,\OOxy(x_3,y_3) \rangle\,
=\,\frac{1}{N_c} D_{12}\,D_{23}\,D_{31}  +  \frac{1}{N_c}\mathbb{T}_{\text{extr}}\,.
\eeq
In order to unpack the correlators from the generating function, one can simply expand back the geometric series:
\beq
\frac{d_{ij}}{1-d_{ij}} = \sum_{a=1}^{\infty}(d_{ij})^a
\eeq
and read-off the individual correlators according to their R-charge (exponents of $d_{ij}$). 

The second term $\mathbb{T}_{extr}$ in the  generating function \eqref{eq:GG3point} represents the contribution of the extremal three-point functions with topology shown in figure. Unlike the fully connected part, the extremal part differs slightly for $U(N_c)$ and $SU(N_{c})$:
\bba
\mathbb{T}_{\text{extr}}^{U(N_c)} &=  D_{12} D_{23} \,+\,  D_{23} D_{31}  + \,D_{31} D_{12} \,,
\\
\mathbb{T}_{\text{extr}}^{SU(N_c)} &= d_{12} d_{23} D_{12} D_{23} \,+\, d_{23} d_{31} D_{23} D_{31}  + d_{31} d_{12}\,D_{31} D_{12}\,.
\end{align}
The role of the extra factors $d_{ij}$ in the $SU(N_c)$ case is to shift the lowest R-charge from $\OOk_1$ to $\OOk_2$. This result could in principle be written exclusively in terms of our effective propagator using: $d_{ij} = \frac{D_{ij}}{1+D_{ij}}$, at the cost of breaking the polynomial structure on $D_{ij}$.

In fact, there is a natural third option for $\mathbb{T}_{\rm extr}$: it can be removed altogether
by a change in operator basis.
Indeed, order by order in  $1/N_c$, it is often natural to
add a multiple of multi-trace operators to single-trace ones, in order to make them mutually orthogonal.  This is called the single-particle basis in refs.~\cite{Aprile:2018efk,Aprile:2020uxk}. In this basis,
by definition, $\mathbb{T}_{\text{extr}}$ vanishes:
\beq\label{eq:SPbasis}
\mathbb{T}_{\text{extr}}^{SP}=0\quad\mbox{where}\quad
\OOk^{SP}_{k} \,\equiv \, \OOk_{k} + \,\sum_{k'} a_{k,k'}\,  \OOk_{k'} \OOk_{k-k'} +\ldots
\eeq

Let us move on to four-point functions. In the free theory these can be similarly
written as a sum over tree-level graphs
labelled by six propagator powers $l_{ij}$ where $1\leq i<j\leq 4$.
Some graphs occur with different symmetry factors; for example, the free correlator with equal weights is:
\bba
G^{\rm free}_{kkkk} &= \sum_{a=1}^{k-1} \left(d_{12}d_{34}\right)^a\left(d_{23}d_{14}\right)^{k-a} + \left(1\leftrightarrow 2\right) + \left(1\leftrightarrow 4\right)\nonumber\\
&\qquad\; +\sum_{a=2}^{k-1}\sum_{b=1}^{a-1} 2\,\left(d_{12}d_{34}\right)^{k-a}\left(d_{13}d_{24}\right)^{a-b}\left(d_{14}d_{23}\right)^b
\end{align}
where the operations  such as $(1\leftrightarrow 2)$ act on the first term by swapping the indexes of $d_{ij}$.  Similar results are known for correlators with different weights, for example in \cite{Caron-Huot:2018kta} (appendix B.1 therein).

\begin{figure}[t]
 \begin{minipage}[h]{1\textwidth}
 \centering 
  \resizebox{3\totalheight}{!}{\includestandalone[width=1\textwidth]{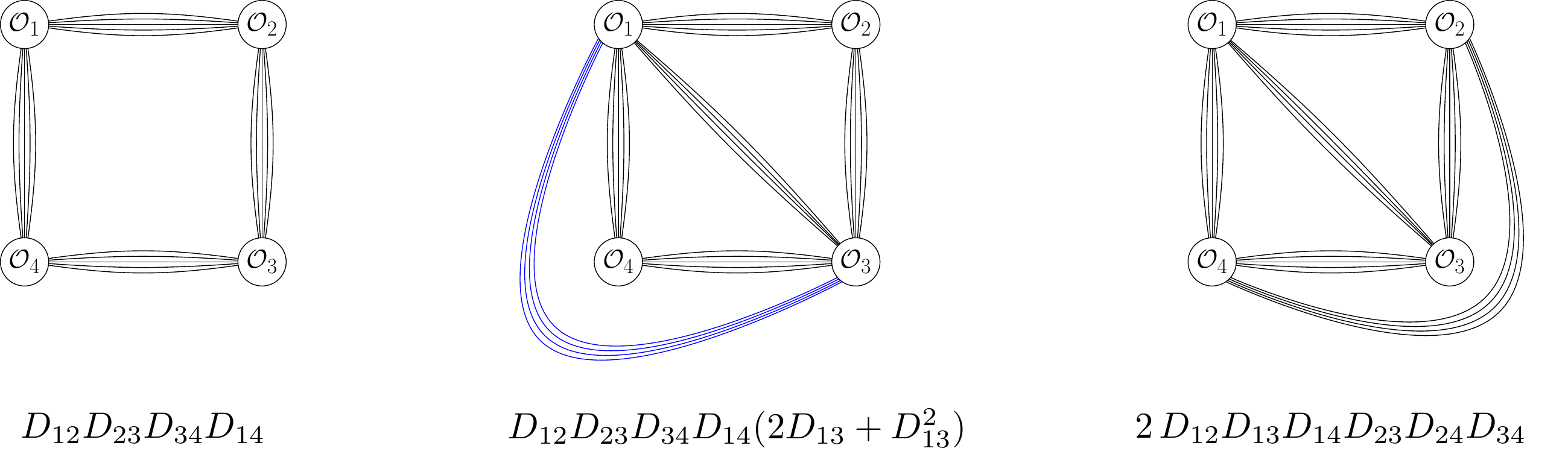}}
\end{minipage}
\caption{Non-extremal four-point topologies, where each operator couples to at least two distinct other operators.  Bundles denote products of one or more Wick contractions.}
\label{fig:PickWick}
\end{figure}

In principle, by summing such formulas, one could compute the generating function:
\beq
G^{\rm free} \,\equiv\, N_c^2\langle \OOxy(x_{1},y_{1})\,\OOxy(x_{2},y_{2})\,\OOxy(x_{3},y_{3})\,\OOxy(x_{4},y_{4})\rangle_c = \sum_{k_{i}\geq 1} G^{free}_{k_1 k_2 k_3 k_4}\,.
\eeq
The easiest way however is to enumerate the possible topologies as shown in fig.~\ref{fig:PickWick}.
Since each graph in each family comes with the same symmetry factor up to a factor
$k_1k_2k_3k_4$ (coming from cyclic permutations of each operator) which cancels out in
the definition of $\mathcal{O}_k$ (see eq.~\eqref{eq:Ok}), each bundle simply yields the ten-dimensional propagator $D_{ij}$. Adding the topologies then gives the generating function of
four-point correlators:
\bba
G^{\rm free} \,
&=\, D_{12}D_{23}D_{34}D_{14}(1+2\,D_{13}+D_{13}^2+2\,D_{24}+D_{24}^2)  + (1\leftrightarrow 2)+(1\leftrightarrow 4) \nonumber\\
&\qquad  + 2\,D_{12}D_{13}D_{14}D_{23}D_{24}D_{34} \,+\,G^{\rm free}_{\rm extr}
\label{eq:GGfree}\,.
\end{align}
The first term, $D_{12}D_{13}D_{34}D_{14}$, generates all graphs with four bundles forming a square with corners (1234) and arbitrary number of propagators on each bundle, see figure \ref{fig:PickWick}. Furthermore, when dressed with the factor:
\beq
2D_{13}+D_{13}^2 = \sum_{a=1}^{\infty} (a+1)\,(d_{13})^a
\eeq
it gives all graphs with extra propagators between operators $\OOk_{1}$ and $\OOk_{3}$. The coefficient $a+1$ on the right-hand side gives the number of ways of splitting the diagonal bundle with $a$ propagators distributed between the inside and outside of the square frame as shown in the figure.

Other terms in \eqref{eq:GGfree} give permutations of the square frame and its diagonals. The last term generates all graphs with all six bundles turned on between the six distinct pairs, see figure. The factor of 2 accounts for the two distinct ways of drawing the diagonals inside or outside the square frame.

\begin{figure}[t]
 \begin{minipage}[h]{1\textwidth}
 \centering 
  \resizebox{2\totalheight}{!}{\includestandalone[width=1\textwidth]{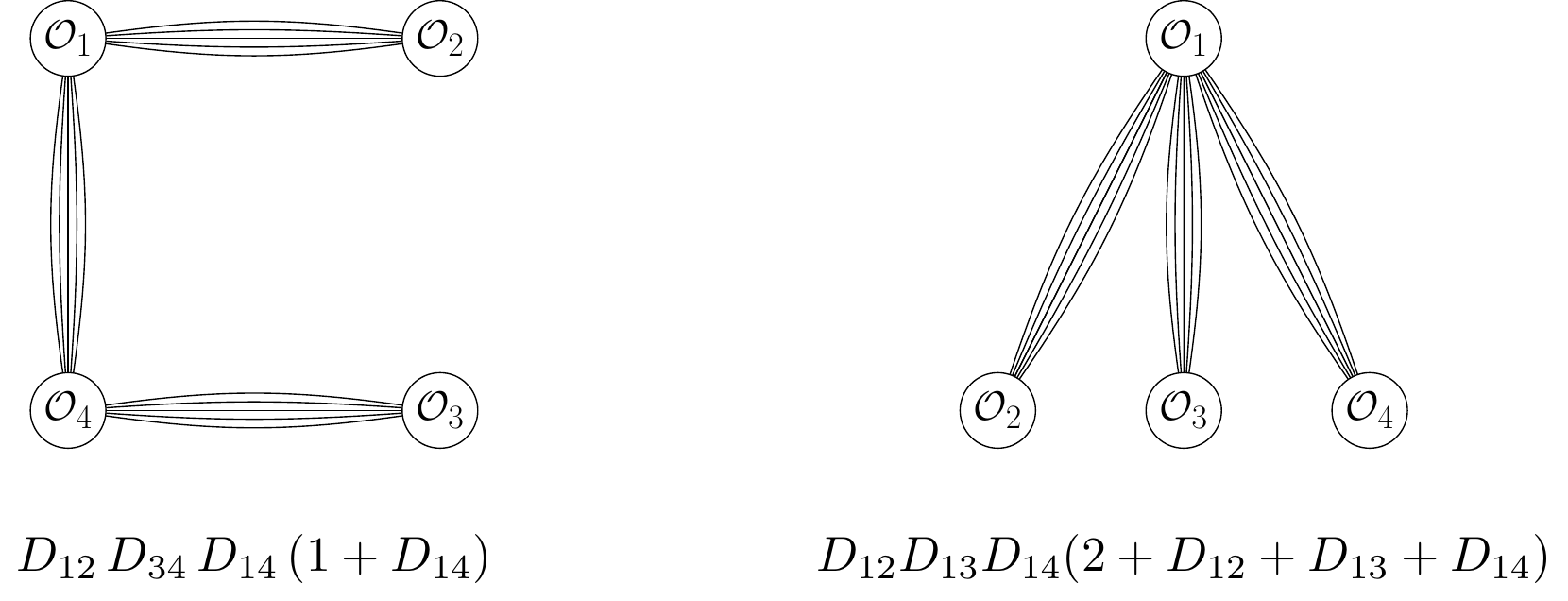}}
 \end{minipage}
 \caption{Extremal topologies. At the bottom the corresponding generating function for the $U(N_c)$ case.}
 \label{fig:PickWickExt}
 \end{figure}
 
Diagrams contributing to the extremal pieces are shown in fig.~\ref{fig:PickWickExt}.
As in the previous three-point example, the ``extremal'' piece vanishes in the single-particle basis:
\beq
G^{{\rm free},SP}_{\rm extr} =0\,.
\eeq
We record for completeness the result in the field theory basis in the $U(N_c)$ case:
\bba\label{eq:GG4extr}
G^{{\rm free},U(N_{c})}_{\rm extr}
&= \,D_{12}\,D_{34}\,D_{14}\,(1+D_{14}) + (\text{11 permutations})\nonumber\\ &\quad+D_{12}\,D_{34}\,D_{14}\,D_{13}\,(2+D_{14}+D_{13})+ (\text{11 permutations})\nonumber\\
&\quad +  D_{12} D_{13} D_{14} \,(2+D_{12} + D_{13} + D_{14})+ (\text{3 permutations})\,.
\end{align}
This generates all graphs which become disconnected after removing one bundle.
For instance, the first and third lines in \eqref{eq:GG4extr} generate the topologies in figure \ref{fig:PickWickExt}.  The $SU(N_c)$ result can be obtained straightforwardly from this
by subtracting terms with only a single power of any given $d_{ij}$. For the first two lines,
this amounts to simple multiplication with extra $d_{ij}$ factors,
but for the third line the result is much less elegant.

At loop level, U(1) factors and extremal structures decouple,
because the U(1) sector is non-interacting and extremal structures would
represent loop corrections to 3- or 2- point functions, which are protected.
Therefore, for loop corrections, we will not need to distinguish $SP$, $U(N_c)$ or $SU(N_c)$ correlators.

\subsection{Hidden symmetry at one-loop}

We saw that geometric series of free theory correlators are concisely captured by using
ten-dimensional propagators $D_{ij}=\frac{y_{ij}^2}{X_{ij}^2}$.
Yet, free correlators do not enjoy ten-dimensional symmetry
because of the $y_{ij}^2$ numerators in $D_{ij}$.
However, we will find that loop integrands, for the \emph{reduced} correlators,
\emph{do} enjoy a full 10D symmetry!

At one loop, the integrand was recorded in eq.~\eqref{one loop}, which we repeat for convenience:
\beq
\mathcal{H}^{(1)}_{k_{1}k_{2}k_{3}k_{4}} =  \frac{1}{\prod\limits_{1\leq i < j \leq 5}x_{ij}^{2}}\,\sum_{\{b_{ij}\} \atop k_{i} =2+\sum_{j} b_{ij}}\, 
\times \prod_{1\leq i<j \leq 4} \left(d_{ij}\right)^{b_{ij}}\,.
\eeq
Summing over $k_i$ one finds independent geometric series in each $b_{ij}$.
Recalling that $d_{ij}=\frac{-y_{ij}^2}{x_{ij}^2}$, this can be written entirely in terms of
ten-dimensional distances!
\beq
\mathcal{H}^{(1)}(X_{ij}^2)=\frac{1}{\prod\limits_{1\leq i < j \leq 5}X_{ij}^{2}}
\quad\text{with}\quad X_{ij}^2 = x_{ij}^2+y_{ij}^2\,. \label{one loop H}
\eeq
For $i\geq 5$ we set $y_i=0$.  We stress that ten-dimensional symmetry holds for
the \emph{reduced} correlator $\mathcal{H}$ but not for $\mathcal{G}$ itself.

The planarity of the one-loop correction \eqref{one loop H} deserves comment---the displayed
product does not correspond to a planar graph!
However, analyzing the factor $R_{1234}$ from eq.~\eqref{R}, one finds that it can be written
as a sum of terms which each contain at least one $X_{ij}^2$ factor. This suffices to make
the product $R_{1234}\mathcal{H}^{(1)}$ planar.
A similar phenomenon was noticed in the integrability context \cite{Fleury:2016ykk}.
At higher loops, $\mathcal{H}^{(\ell)}$ itself will be a sum of planar integrands.

In this example, the 10-dimensional symmetry could be turned around to
recover the generic correlator from the simplest case:
\beq
\mathcal{H}^{(1)}(x,y)= \mathcal{H}^{(1)}_{2222}(x)\Big{|}_{x_{ij}^2\to x_{ij}^2 + y_{ij}^2}\,.
\eeq
This is a powerful observation which will be put to use in the next subsections.

\subsection{Conjectured 10-dimensional symmetry for higher-loop integrands}
\label{ssec:conjecture}

We are ready to state our main conjecture: that the (reduced) integrands at each loop
order derive from a generating function $\mathcal{H}^{(\ell)}(X_{ij}^2)$
which enjoys full ten-dimensional conformal symmetry, and
full permutation symmetry in $\ell+4$ coordinates.

A given correlator is obtained simply by extracting the coefficient of the correct power of $y_i$,
explictly:
\beq\label{eq:Hprediction}
\boxed{
\mathcal{H}^{(\ell)}_{k_{1} k_{2} k_{3} k_{4}}(x_{ij}^2,y_{ij}^2)
=
\mbox{coefficient of $\left(\prod_{i=1}^4 \beta_i^{k_i-2}\right)$ in } \mathcal{H}^{(\ell)}(X_{ij}^2)\,
\big{|}_{y_{ij}^2 \to \beta_{i}\beta_{j}\,y^2_{ij}} \,,}
\eeq
where we recall that $X_{ij}^2=x_{ij}^2 + y_{ij}^2$.
The same dictionary was introduced previously in the supergravity limit \cite{Caron-Huot:2018kta}.

In particular this implies that the truncation from 10D to 4D of the generating function is 
the integrand of the simplest multiplet:
\beq\label{eq:reduction4D}
\lim_{X_{ij}^2\to x_{ij}^2}\mathcal{H}^{(\ell)}(X_{ij}^2) = \mathcal{H}_{2222}(x_{ij}^2)\,.
\eeq
Conversely, the conjecture implies that the $\mathcal{H}_{2222}^{(\ell)}$ integrand
determines all other correlators modulo functions which vanish in $D=4$, that is, vanishing
Gram determinants:
\beq\label{inverse}
\mathcal{H}^{(\ell)}(X_{ij}^2) = \mathcal{H}^{(\ell)}_{2222}(x_{ij}^2)\Big|_{x_{ij}^2\mapsto X_{ij}^2}
+ \mbox{(Gram determinants)}\,.
\eeq
As will be discussed in section \ref{sec:Gram}, no planar, permutation symmetric,
conformally invariant Gram determinant identity
exist below 8 loops. Equations \eqref{eq:Hprediction} and \eqref{inverse} thus unambiguously predicts
all R-charged correlators up to at least 7-loops.

\subsection{Two and three-loop correlators}
\label{sec:Integrated23}

Let us illustrate the preceding conjecture at low loop orders.
The two-loop generating function obtained from \eqref{eq:H2twoloop}
using the replacement \eqref{inverse} is simply:
\beq
\mathcal{H}^{(2)} \,=\,\frac{X_{12}^2 X_{34}^2 X_{56}^2 \,+\, \text{14 permutations}}{\prod\limits_{1\leq i < j \leq 6}X_{ij}^2} \label{eq:2loopH}\,.
\eeq
To relate this to usual correlators, different permutations of the 6 coordinates will be treated differently.
We recall that $y_{i}=0$ for the Lagrangian insertions ($i\geq 5$), so that
for example $X_{i5}^2=x_{i5}^2$.  For the other distances we use the relation $X_{ij}^2=(1-d_{ij})x_{ij}^2$:
\bba\label{eq:2generatingH}
\mathcal{H}^{(2)} \,&=\, \frac{1}{{\color{blue}\prod\limits_{1\leq i< j\leq 4}(1-d_{ij})}\prod\limits_{1\leq i < j \leq 6}x_{ij}^2}
(1+\sigma+\sigma^2)
\\ \nonumber
&\bigg{(} {\color{blue}(1-d_{12})(1-d_{34})}x_{12}^2x_{34}^2x_{56}^2 
+\left[{\color{blue}(1-d_{34})}x_{34}^2 x_{16}^2 x_{25}^2 +{\color{blue}(1-d_{12})}x_{12}^2 x_{45}^2 x_{36}^2+ (5\leftrightarrow 6)\right] \bigg{)}
\end{align}
where $\sigma$ denotes a cyclic rotation of $(234)$.
The correlators $\mathcal{F}$ defined in eq.~\eqref{eq:fromHtoF} may then be read off simply by expanding
the geometric series $\frac{1}{1-d_{ij}}= \sum\limits_{m=0}^{\infty}(d_{ij})^m$ and collecting all terms with common powers $(d_{ij})^{b_{ij}}$:
\beq
\mathcal{H}^{(\ell)} = \sum_{b_{ij}\geq 0} \mathcal{F}^{(\ell)}_{\{b_{ij}\}}\,\prod_{1\leq i < j \leq 4}(d_{ij})^{b_{ij}}\,.
\eeq
Some of the simplest are:
\bba
\mathcal{F}^{(2)}_{\{0,0,1,1,1,1\}} &= \left(\frac{x_{13}^2}{(x_{15}^2 x_{25}^2 x_{35}^2) x_{56}^2 (x_{16}^2  x_{36}^2 x_{46}^2)} +(x_5\leftrightarrow x_6)\right) \, + (x_2\leftrightarrow x_3)\,,\\ 
\mathcal{F}^{(2)}_{\{1,0,1,1,1,1\}} &= \frac{x_{13}^2}{(x_{15}^2 x_{25}^2 x_{35}^2) x_{56}^2 (x_{16}^2  x_{36}^2 x_{46}^2)} +(x_5\leftrightarrow x_6)\,,\\
\mathcal{F}^{(2)}_{\{1,1,1,1,1,1\}} &= 0\,.
\end{align}
and similarly $\mathcal{F}^{(2)}_{\{0,\beta_{1},\beta_{2},\beta_{3},\beta_{4},\beta_{5}\}} =\mathcal{F}^{(2)}_{\{0,1,1,1,1,1\}}$ when all $\beta_{i} \geq 1$.
We see also that the one-loop-squared integral only appears in $\mathcal{F}$ with two or more zeros, for instance:
\beq
\mathcal{F}^{(2)}_{\{1,0,1,1,0,1\}} \,=\,\frac{x_{13}^{2}\,x_{24}^2}{\prod\limits_{a=5,6}\,\prod\limits_{ i=1,2,3,4}x_{ia}^2}\,+\,\left[\frac{1}{(x_{15}^2 x_{25}^2) x_{56}^2 ( x_{36}^2 x_{46}^2)}\left(\frac{x_{13}^2}{x_{35}^2\,x_{16}^2} +\frac{x_{24}^2}{ x_{45}^2 x_{26}^2}\right)+(x_{5}\leftrightarrow x_{6})\right].
\eeq
Moving to integrated expressions, since the $d_{ij}$ factors commute with integration over $x_{i\geq 5}$, a generating function of R-charged correlators immediately follows:
\beq\label{eq:GGfromHH}
G^{(\ell)}
\equiv \, \sum_{k_{i}\geq 2}G^{(\ell)}_{k_{1}k_{2}k_{3}k_{4}}
= \frac{(-g^2)^\ell}{\ell!} R_{1234}\left(2\,x_{12}^2x_{13}^2 x_{14}^2x_{23}^2x_{24}^2x_{34}^2\right)\,\int \frac{dx_{5}^4}{\pi^2}\cdots \frac{dx_{4+n}^4}{\pi^2}\,\,\mathcal{H}^{(\ell)}\,.
\eeq
At one and two-loops, the formulas recorded above give
\bba
G^{(1)} &=-2g^2 R_{1234}\, g_{1234}\,\prod_{1\leq i < j\leq 4}\frac{1}{1-d_{ij}}\,,\\
G^{(2)} &= 2g^4 R_{1234}\bigg{(}{\color{blue}c^1_h}\,h_{12;34} +c^2_h\,h_{13;24}+c^3_h\,h_{14;23}
+ \frac{1}{2}\big{(}{\color{blue}c^{1}_{gg}}\,x_{12}^2 x_{34}^2+c^{2}_{gg}\,x_{13}^2 x_{24}^2+c^{3}_{gg}\,x_{14}^2 x_{23}^2\big{)} \left[g_{1234}\right]^2\bigg{)}
\end{align}
where $g$ and $h$ are the one- and two-loop ladders defined in eq.~\eqref{eq:1-2loopsCI},
and the independent coefficients are:
\beq
{\color{blue}c^1_{h}}=\frac{(1-d_{12})+(1-d_{34})}{\prod\limits_{1\leq i<j \leq 4}(1-d_{ij})}\quad\text{and}\quad {\color{blue}c^{1}_{gg}}= \frac{(1-d_{12})(1-d_{34})}{\prod\limits_{1\leq i<j \leq 4}(1-d_{ij})}\,.
\eeq
Through these manipulations, we have shown that eq.~\eqref{eq:2loopH} fully reproduces table \ref{tab:2loops}!

At three-loops, all R-charged correlators can now be obtained similarly by uplifting \eqref{eq:H2threeloop},
which yields:
\bba\label{eq:all3loop}
G^{(3)}&=-2g^6\,R_{1234} \bigg{(}{\color{blue}c^{1}_{gh}}(g\times h)_{12;34}+c^{2}_{gh}(g\times h)_{13;24}+c^{3}_{gh}(g\times h)_{14;23}\nonumber\\
&+ {\color{blue}c^{1}_{L}}L_{12;34}+ c^{2}_{L}L_{13;24}+ c^{3}_{L}L_{14;23}
+ {\color{blue}c^{1}_{E}}E_{12;34}+ c^{2}_{E}E_{13;24}+ c^{3}_{E}E_{14;23}\nonumber\\
&+\frac{1}{2}({\color{blue}c^{1}_{H}}+{\color{blue}c^{2}_{H}}\,1/v)H_{12;34}+ +\frac{1}{2}(c^{3}_{H}+c^{4}_{H}\,u/v)H_{13;24}+\frac{1}{2}(c^{5}_{H}+c^{6}_{H}\,u)H_{14;23}\bigg{)}
\end{align}
with the conformal integrals given in appendix~\ref{app:3loops} and coefficients (with
$P_{4}\equiv\prod\limits_{1\leq i<j\leq 4}(1-d_{ij})$):
\bba
c^{1}_{gh}\,&=\,  (1-d_{12})(1-d_{34})[(1-d_{12})+(1-d_{34})]/P_4\,,\nonumber \\
c^{1}_{L}\,&=\,   [(1-d_{12})^2  +(1-d_{34})^2+ 2(1-d_{12})+2(1-d_{34})]/P_4\,,\nonumber\\
c^{1}_{E}\,&=\,   [(1-d_{14})+(1-d_{23})][(1-d_{13})+(1-d_{24})]/P_4\,,\nonumber\\
c^{1}_{H} \,&=\,   (1-d_{14})(1-d_{23})[(1-d_{12})+(1-d_{34})]/P_4\,,\nonumber\\
c^{2}_{H}  \,&=\,   (1-d_{13})(1-d_{24})[(1-d_{12})+(1-d_{34})]/P_4\,.
\end{align}
This is again in complete agreement with refs.~\cite{Chicherin:2015edu,Chicherin:2018avq}.

\subsection{Absence of Gram determinant ambiguities with 7 loops or less}
\label{sec:Gram}

The conjectured all-order symmetry \eqref{eq:Hprediction} allows the R-charged correlators to be directly
uplifted from the better-known integrands $\mathcal{H}_{2222}$,
modulo the Gram ambiguities in eq.~\eqref{inverse}.  These ambiguities are restricted by the full permutation and conformal symmetries, and the condition that they originate from planar diagrams.
We find that it is very difficult to write \emph{planar} ambiguities.

An ambiguity in uplifting $\mathcal{H}_{2222}$ can be characterized
as a sum of planar conformal diagrams which vanishes when restricted to $d=4$:
\beq
 \mbox{(Gram ambiguity)} \equiv \sum c_i \mathcal{I}_i(X_1,\ldots X_{4+\ell}) \mbox{ which vanishes when $X_i\in \mathbb{R}^4$}\,
\eeq
where the $c_i$ are pure numerical factors, and the $\mathcal{I}_i$ are permutation-invariant conformal integrands in $\ell+4$ arguments.
Such functions were enumerated in ref.~\cite{Eden:2012tu}.
By evaluating the integrands numerically at several
random points in $\left(\mathbb{R}^4\right)^{\ell+4}$ and computing the rank of the resulting matrix, we can then easily determine if a nontrivial combination $c_i$ exists.  In this way we proved
that planar Gram determinants do not exist at least up to seven loops!  That is,
the conjecture \eqref{eq:Hprediction} implies that:
\beq \label{inverse7}
\mathcal{H}^{(\ell)}(X_{ij}^2) = \mathcal{H}^{(\ell)}_{2222}(x_{ij}^2)\Big|_{x_{ij}^2\mapsto X_{ij}^2}\qquad\mbox{for } \ell\leq 7\,.
\eeq
The integrand $\mathcal{H}_{2222}$ was given up to 10-loops in \cite{Bourjaily:2016evz}.
Up to 5-loops, we have verified that the prediction in eq.~\eqref{inverse7} is in perfect agreement
with the results of \cite{Chicherin:2015edu,Chicherin:2018avq}.
At 6 and 7 loops, the formula makes new predictions.
At 8 loops, we do not know if a non-trivial Gram ambiguity actually exists, or not.

\section{Octagons as scattering amplitudes on the Coulomb branch}
\label{sec:NewDuality}

Correlators with large $R$-charge have been widely studied. They are a kind of
``simplest four-point functions'' which can be computed at finite-coupling using integrability.
Using properties of the generating functional $\mathcal{H}$, in this section we show that
they are closely related to scattering amplitudes on the Coulomb branch.
This will be understood as a generalization of the correlator/amplitude duality.

We first discuss these relations at the integrand level. We will then observe that
at generic location on the Coulomb branch, all integrals converge so the duality
extends to integrated expression.

\begin{figure}[t]
\centering
 \resizebox{2.25\totalheight}{!}{\includestandalone[width=.8\textwidth]{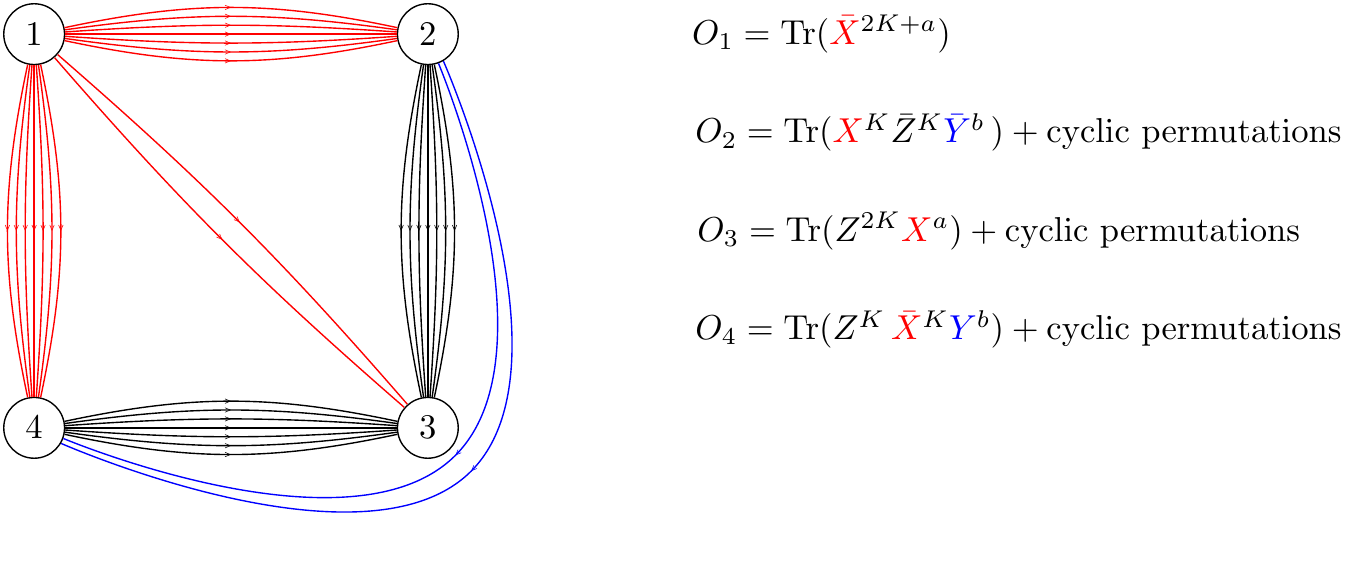}}
 \caption{A family of simple four-point functions. The polarization choices ensure that
 a single Wick contraction exists in the field theory.    $K$ is assumed to be larger than loop order so as to make the four exterior bundles effectively uncrossable.
 }
\label{fig:Simplest}
\end{figure}

Octagons are defined in the literature by considering correlators with large R-charge so as to create four fat edges.  This can be achieved using the polarizations shown in fig.~\ref{fig:Simplest}. These polarized correlators can be extracted from the unpolarized ones by identifying the R-charge structure:
\bba
 \lim_{K\to\infty}\langle&\OOk_{2K+a}(x_{1}) \OOk_{2K+b}(x_{2}) \OOk_{2K+a}(x_{3})\OOk_{2K+b}(x_{4})\rangle\nonumber\\ 
 &=(d_{12}d_{23}d_{34}d_{41})^K\,(d_{13})^a (d_{24})^b\; S_{a,b} +\cdots
\end{align}
where the dots stand for other R-charge structures which are projected out by our choice of polarizations.

The ``simplest correlator'' $S_{a,b}$ is defined more precisely by taking the limit $K\to\infty$,
which implies a factorization into two squares, referred to as octagons.
In perturbation theory, this factorization is a consequence of the suppression of all  planar Feymann diagrams that cross the square perimeter formed by the bridges of length $K$. Such a planar Feynman graph would contribute at the order $\sim(g^2)^K$ and hence can be neglected in the large charge limit. 
 

Because $S$ factorizes, it is natural to take its square root. This defines the \emph{octagon}:
\beq
\mathbb{O}\times \mathbb{O}\coloneqq \sum_{a,b\geq 0} d_{13}^a d_{24}^b S_{a,b}\,.
\eeq
Let us express the integrand for $S_{a,b}$ in terms of the correlators $\mathcal{F}_{a,b}$ which we discussed above. Defining integrands similarly to eq.~\eqref{eq:Gcorrelator},
\beq
 S_{a,b} = \sum_\ell \frac{(-g^2)^\ell}{\ell!} \int \frac{d^4x_5}{\pi^2}\cdots\frac{d^4x_{4+\ell}}{\pi^2}
 \mathcal{S}^{(\ell)}_{a,b}
\eeq
and comparing with eqs.~\eqref{eq:susyward} and \eqref{eq:fromHtoF}, we get:
\bba \label{Sab}
\mathcal{S}^{(\ell)}_{a,b} = 2 x_{13}^2 x_{24}^2\left(F^{(\ell)}_{a,b}-2F^{(\ell)}_{a-1,b-1}+F^{(\ell)}_{a-2,b-2}\right)\quad \text{with }F^{(\ell)}_{a,b}\equiv\, \mathcal{F}^{(\ell)}_{\{\infty,a,\infty,\infty,b,\infty\}}\prod_{1\leq i<j\leq 4}x_{ij}^2\,.
\end{align}
First we notice that all theses R-charge structures contributing to the simplest correlators have
$\infty$ exponents for the fat bridges $d_{12}, d_{23}, d_{34}, d_{41}$.
Furthermore, by analyzing the generating function, we notice that these structures only received contributions from the terms in $\mathbb{G}$ which contain all the four poles $(1-d_{i,i+1})$,
since only the geometric series of $\frac{1}{1-d_{i,i+1}}$ can give arbitrarily high exponents.
We conclude that the octagon (squared) can be extracted at once by applying the null limit:
\beq
\text{10D null-square limit:}\quad d_{i,i+1}\to 1   \,\equiv\,  X^2_{i,i+1}\to 0  \,\equiv\, y_{ij}^{2} \to -x_{ij}^{2}    \qquad\qquad\qquad
\eeq
to the generating function of all correlators G:
\bba\label{eq:fromGtoO2}
 \mathbb{O}\times \mathbb{O}\big|_{\rm integrand}&=
 \,
 2\,x_{13}^4x_{24}^4(1-d_{13}d_{24})^2 \Res\limits_{X_{i,i+1}^2=0} \mathcal{H}\,
\\
&= 
\Res\limits_{d_{i,i+1}=1} G\big{|}_{\text{integrand}}
\end{align}
where a residue is taken for each of the four propagators defining the 10D null square.
The prefactor on the first line
accounts for the three terms in eq.~\eqref{Sab} and can be understood
from $R_{1234}\to x_{13}^2x_{24}^2(1-d_{13}d_{24})^2$ in eq.~\eqref{R},
which also explains why it cancels when we go back to the (unreduced) correlator.
In the free theory, we find from eq.~\eqref{eq:GGfree}:
\beq
 \mathbb{O}^{\rm free} = \frac{1-d_{13}d_{24}}{(1-d_{23})(1-d_{14})}\,.
\eeq

\begin{figure}[t]
\begin{minipage}[h]{1\textwidth}
 \centering 
  \resizebox{5.5\totalheight}{!}{\includestandalone[width=.5\textwidth]{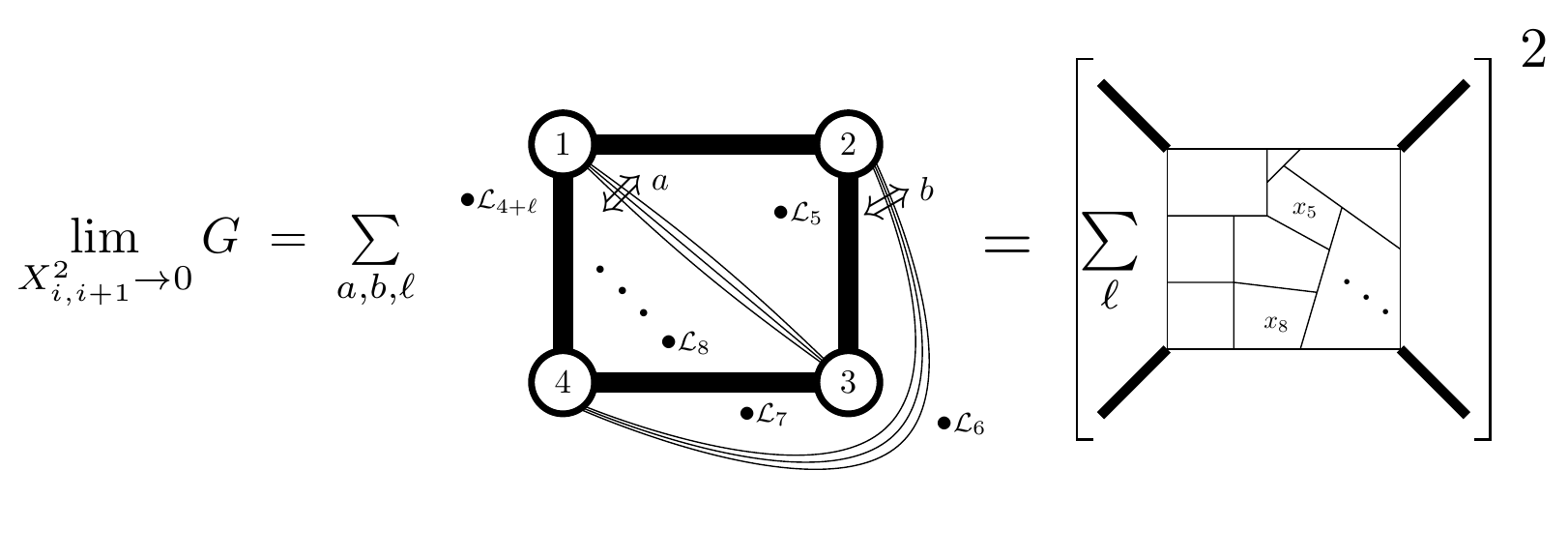}}
 \end{minipage}
 \caption{Extended correlator-amplitude duality: The generating function of all simplest correlators, obtained as the 10D null limit of $G$, is equal to the square of a four-point  amplitude of massive particles in the Coulomb branch of $\mathcal{N}=4$ SYM. On the left we sum over all diagonal bridges $a$ and $b$ and Lagrangian insertions; their coordinates $x$ become region momenta on the right.
The thick lines in the square perimeter represent the infinite $R$-charge. On the right the thick lines at the corners represent the four massive external particles and the thin lines are massless internal propagators.}
 \label{fig:CAduality}
 \end{figure}

Let us interpret this formula.  When the $y$'s vanish,
the residue in eq.~\eqref{eq:fromGtoO2} is similar to that entering
the familiar (OPE-type) relation between the null limit of a correlation function and a Wilson loop (here, integrand)
\cite{Alday:2010zy},
which is itself dual to a gluon scattering amplitude \cite{Alday:2007hr,Berkovits:2008ic,Beisert:2008iq,Bern:2008ap,Drummond:2008aq,
Mason:2010yk,CaronHuot:2010ek,Eden:2010zz}.
While it is presently not obvious how to attach the $y$ variables onto a Wilson loop,
the amplitude side of that duality suggests  a clear way forward.
Namely, the $y_i$ coordinates admit a natural physical interpretation as
vacuum expectation values for the scalar fields of the $\mathcal{N}=4$ theory \cite{Alday:2009zm}.
Some of the gluons become massive through the Higgs mechanism, and
the amplitude (integrands) for these massive $W$ bosons are already known to enjoy
a ten-dimensional dual conformal symmetry \cite{CaronHuot:2010rj}.
This suggests viewing the 10D null limit in eq.~\eqref{eq:fromGtoO2} as
a scattering amplitude of $W$ bosons on the Coulomb branch:
\beq\label{eq:O=M}
\boxed{\frac{\mathbb{O}(x,y)}{\mathbb{O}^{\text{free}}(x,y)} =  M(x,y)}
\eeq
where now $M$ is a
massive four-particle amplitude in the Coulomb branch (normalized so that $M^{\text{free}}=1$),
with external momenta and masses satisfying:
\beq
p_{i}^\mu \equiv x^\mu_{i,i+1}\quad \text{and}\quad m_{i}^2 \equiv y_{i,i+1}^2\,.
\eeq
The 10D null-limit $X_{ij}^2\to 0$ is then simply the on-shell condition on the Coulomb branch: $p_i^2+m_i^2=0$. The duality is depicted in fig.~\ref{fig:CAduality}.

In other words, we propose that the octagon is precisely a scattering amplitude on the Coulomb branch.

The octagons defined in this paper span only a restricted subset of the Coulomb branch,
where each Higgs vev satisfies $y_i^2=0$.
(Nonzero external masses $y_{i,i+1}^2\neq 0$ can still be realized by allowing the $y_i$ to be complex 6-vectors.)
This suggests the existence of a more general, truly ten-dimensional object
which relaxes the constraint $y_i^2=0$, but this seems difficult to capture using local operators in the field theory
and is beyond the scope of this paper.

\subsection{Example amplitudes on the restricted Coulomb branch}

\begin{figure}[t]
 \begin{minipage}[h]{1\textwidth}
 \centering 
  \resizebox{2\totalheight}{!}{\includestandalone[width=1\textwidth]{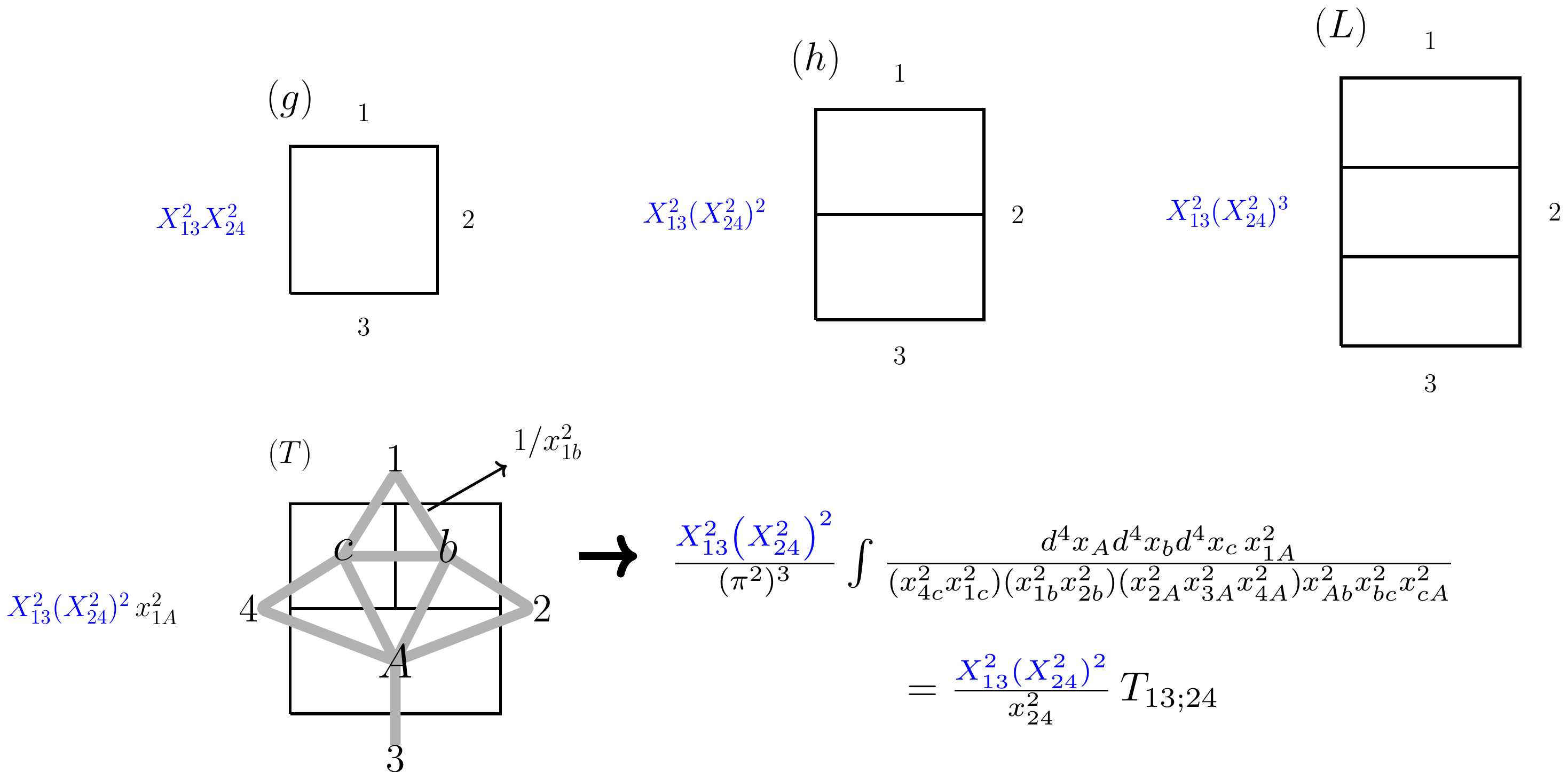}}
 \end{minipage}
 \caption{Conformal integrals in the four-particle amplitude integrand up to 3 loops. The first line shows
 one-loop, two-loop and three-loop ladder integrals in momentum space representations. 
The last line illustrates, for the three-loop tennis-court integral,
the planar graph duality which relates the momentum and position space representations.
 \label{fig:3loopAmplitude}}
\end{figure}

Let us first exemplify these relations at the integrand level by present them up to three loops.
We start with the integrand for stress tensors in \cref{eq:H2oneloop,eq:H2twoloop,eq:H2threeloop}, then uplift to the generating function $G$ using \eqref{inverse}, \eqref{eq:GGfromHH} and take the $10D$ null-limit 
to get $\mathbb{O}$ or equivalently the amplitude
\eqref{eq:fromGtoO2}:
\bba\label{eq:M3}
M&= 1-g^2\,{X_{13}^2 X_{24}^2}\,g_{1234} \nonumber \\
&\qquad+g^4\left[\frac{{X_{13}^2 (X_{24}^2)^2}}{x_{24}^2}\,h_{13;24} +  \frac{{(X_{13}^2)^2 X_{24}^2}}{x_{13}^2}\,h_{24;13}\right] \nonumber \\
&\qquad-g^{6}\,\left[\frac{{X_{13}^2 (X_{24}^2)^3}}{(x_{24}^2)^2}\,L_{13;24} + \frac{{(X_{13}^2)^3 X_{24}^2}}{(x_{13}^2)^2} \,L_{24;13}+\frac{{X_{13}^2 (X_{24}^2)^2}}{x_{24}^2} \,(T_{13;24}+T_{13;42})\right. \nonumber\\
&\qquad\qquad\quad\left.+  \frac{{(X_{13}^2)^2 X_{24}^2}}{x_{13}^2}\,(T_{24;13}+T_{24;31})\right]\,+\,\mathcal{O}(g^8) \,.
\end{align}
The square root of $\mathbb{O}\times \mathbb{O}$ was computed directly at integrand level by dropping
graphs with vertices on the ``reverse side'' of the square.
The dual graphs corresponding to the integrand in eq.~\eqref{eq:M3} are shown in fig.~\ref{fig:3loopAmplitude}.
The extra factors $x_{ij}^2$ in the denominators are there for consistency with the normalization of the ladders ($g$ and $h$)  in \eqref{eq:1-2loopsCI} and the 3-loop ladder $(L)$ and tennis-court $(T)$ in \eqref{eq:3loopsCI}.

The above integrand agrees precisely with that entering the gluon scattering amplitude,
with precisely the same coefficients \cite{Bern:1997nh}.
The fact that the correct Coulomb branch integrand is obtained from the massless case
by simply making all distances ten-dimensional (ie. $x_{13}^2\mapsto X_{13}^2\equiv x_{13}^2+y_{13}^2$,
$x_{24}^2\mapsto X_{24}^2$ in the numerator) is also in agreement with existing results in the scattering amplitudes context \cite{Alday:2009zm}.
Within the restricted Coulomb branch on which the octagon is defined, only those two invariants have a nontrivial 10D component.

\begin{figure}[t]
\centering
\includegraphics[scale=1]{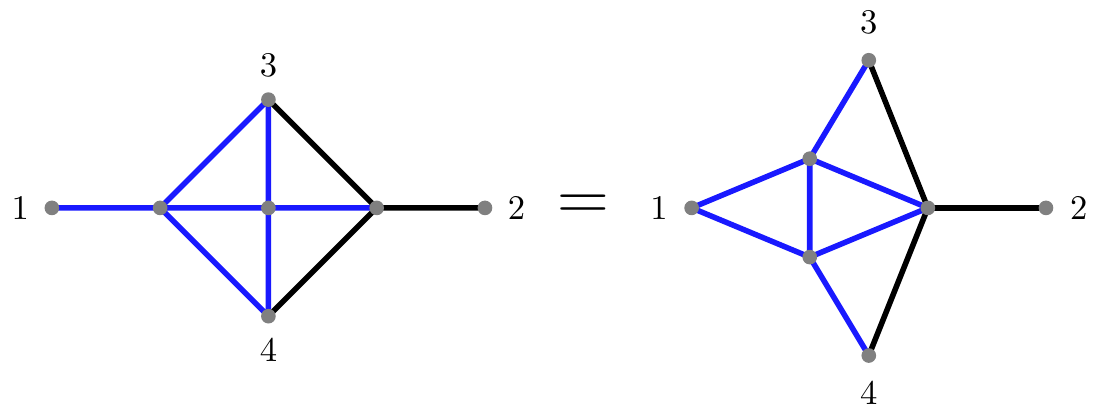}
\caption{Conformal magic identity: the three-loop ladder (left panel) and the tennis court (right panel) can be shown to be identical off-shell by using the conformal invariance of the highlighted two-loop ladder sub-integral.}
\label{fig:LadderVsCourt}
\end{figure}

On the restricted Coulomb branch, the \emph{external} masses $-x_{i,i+1}^2=y_{i,i{+}1}^2$ are
still generically nonvanishing, which makes the amplitude effectively off-shell from the four-dimensional perspective.  This has an important implication: there are no divergences and
conformal symmetry of integrated octagons is unbroken (from the amplitude perspective,
this is called dual conformal symmetry). In particular, the conformal ``magic'' identities of ref.~\cite{Drummond:2006rz} hold exactly on our setup!  For example (see fig.~\ref{fig:LadderVsCourt}):
\beq
T_{13;24} = L_{13;24} = L_{24;13}\,.
\eeq
The integrated form of the three-loop gluon scattering amplitude on the restricted Coulomb branch amplitude can thus be written very simply:
\bba\label{eq:M3loops}
M&=1-g^{2}\,(1-d_{13})(1-d_{24})\,x_{13}^2 x_{24}^2\,g_{1234} \nonumber \\
&\qquad +g^{4}\,(1-d_{13})(1-d_{24})\left(2-d_{13}-d_{24}\right)x_{13}^2 x_{24}^2\,h_{13;24} \nonumber \\
&\qquad -g^{6}\,(1-d_{13})(1-d_{24})\left[(1-d_{13})^2+(1-d_{24})^2+2(2-d_{13}-d_{24})\right]x_{13}^2 x_{24}^2\,L_{13;24}\nonumber\\
&\qquad+\mathcal{O}(g^8)\,.
\end{align}
These functions are given in \cref{eq:1-2loopsCI,eq:3loopsCI}.
The integrated 10D null octagon is identical up to the factor
$\mathbb{O}^{\rm free}=\frac{1-d_{13}d_{24}}{(1-d_{13})(1-d_{24})}$ (see eq.~\eqref{eq:O=M}).

At higher loops, new, harder integrals appear.  On the other hand, the octagon was calculated exactly
using integrability.  In the next section we leverage these results to predict new scattering amplitudes
and Feynman integrals.

\subsection{Comments on the four-dimensional massless limit}
\label{4d massless}

\begin{figure}[t]
\begin{minipage}[h]{1\textwidth}
 \centering 
  \resizebox{1.6\totalheight}{!}{\includestandalone[width=1\textwidth]{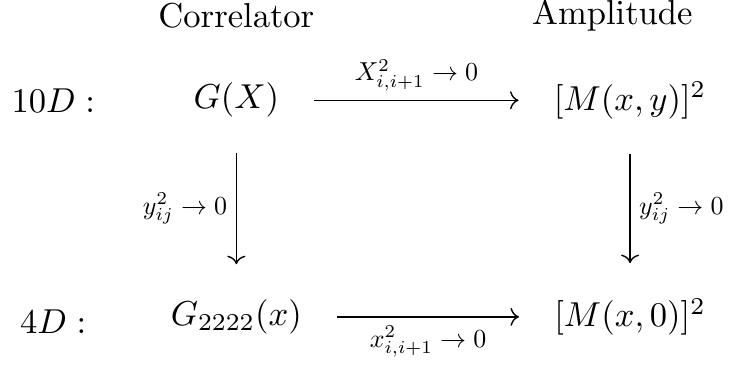}}
 \end{minipage}
 \caption{Relationship between the massive (top) and massless (bottom) correlator-amplitude duality.
 All arrows commute at integrand level.  The different massless limits are however not equivalent at the
 integrated level.}
 \label{fig:CAduality1}
 \end{figure}

In fig.~\ref{fig:CAduality1} we contrast and relate our new duality between the generating function
$G$ and the coulomb branch amplitude, \cref{eq:fromGtoO2,eq:O=M}, with the well-known duality between the correlator of stress tensors and the four-gluon amplitude.
At the integrand level, that duality takes formally the same form \cite{Eden:2010ce}
\beq
 \lim_{x_{i,i+1}^2\to 0} \frac{\mathcal{G}_{2222}(x_i)}{\mathcal{G}_{2222}^{\rm free}} =
M(x_i)^2 \qquad (y=0)\,.
\eeq
In the massless case, however, the loop integrals diverge and
as a result the relation at integrated level takes on a more complicated form:
\beq\label{eq:4Dduality}
 \lim_{x_{i,i+1}^2\to 0} \frac{G_{2222}(x_{i})}{G_{2222}^{\rm free}} \,=\,
 \lim_{\varepsilon\to 0}f(\{\mu_{\rm IR}^2x_{i,i+1}^2\},x_{13}^2,x_{24}^2,\varepsilon)
 \left[M(p_{i}) \right]^{2}\qquad (y=0)
\eeq
where $f$ is a universal function which converts the different regularizations.
For example, if dimensional regularization is used on the scattering amplitude side,
the relation to the $x_{i,i+1}^2\to 0$ limit of the correlator was worked out in refs.~\cite{Alday:2010zy,Eden:2010zz}.

An interesting feature of our novel correlator/amplitude duality \eqref{eq:fromGtoO2} is
that, since the integrals are finite (as long as all $y_{i,i+1}^2\neq 0$), it translates directly to the integrated level!
This is what we have used in eq.~\eqref{eq:M3loops} above.

\begin{figure}[t]
 \begin{minipage}[h]{1\textwidth}
 \centering 
  \resizebox{3\totalheight}{!}{\includestandalone[width=1\textwidth]{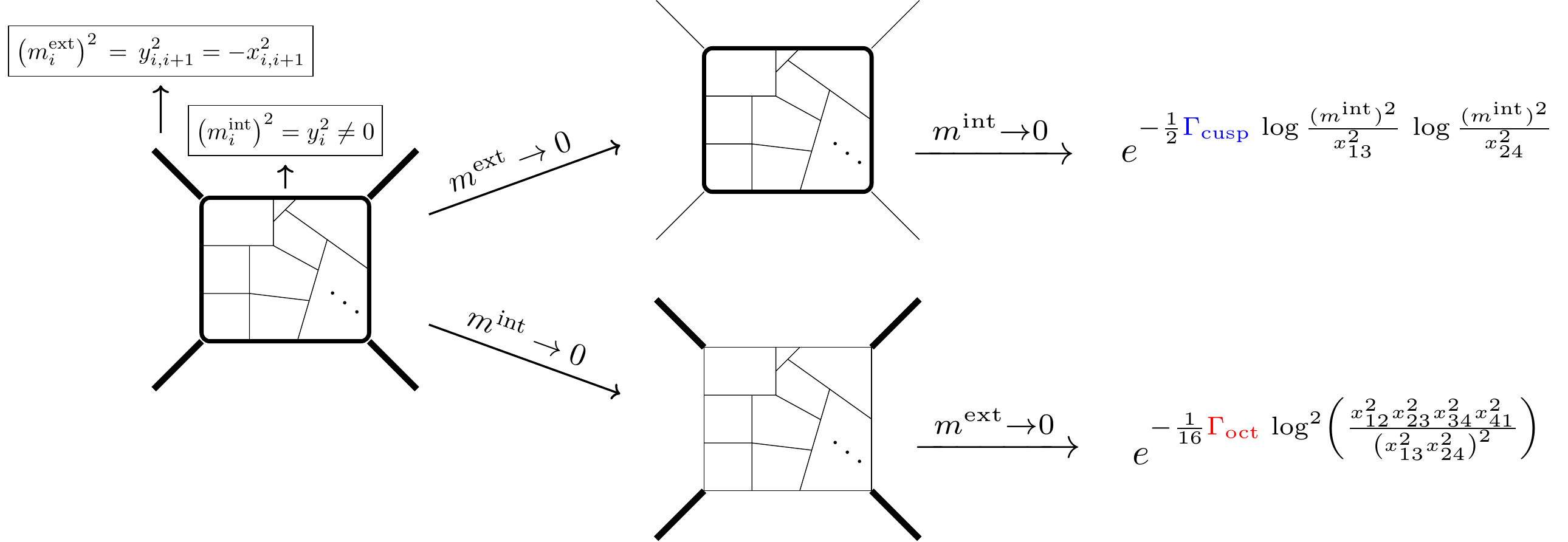}}
 \end{minipage}
 \caption{Different massless limits of Coulomb branch amplitudes
 lead to very different exponential behavior. Both objects in the middle column are infrared-safe.}
 \label{fig:MasslessLimits}
 \end{figure}

As an alternative to dimensional regularization, regularization of scattering amplitudes
using the Coulomb branch has been studied starting from \cite{Alday:2009zm}.
In the simplest setup, one turns on a vev to only one of the six scalars, in such a way as to give masses to the outermost internal propagators of the amplitude as represented in the top of fig.~\ref{fig:MasslessLimits}.
In this limit, one finds a mass-regularized version of the BDS Ansatz \cite{Bern:2005iz,Alday:2009zm}:
\bba
 \log M(x,0) &=- \frac{\Gamma_{\rm cusp}(g)}{4}\left[2 \log\left(\frac{m^2}{x_{13}^2}\right)\,\log\left(\frac{m^2}{x_{24}^2}\right)-\pi^2\right] \nonumber\\
 &\qquad\quad+ \tilde{\mathcal{G}}(g)\left[\log\left(\frac{m^2}{x_{13}^2}\right)+\log\left(\frac{m^2}{x_{24}^2}\right)\right] + \tilde{c}(g) + \OOk(m^2)\,,
 \end{align}
where
\beq
 \Gamma_{\rm cusp}(g)= 4 g^2 -8\zeta_{2}\,g^4 + 88\zeta_{4}\,g^6 -(876\zeta_{6}+32\zeta_{3}^2)\,g^8 + \OOk(g^{10})\,.
\eeq
These kinematics however do not overlap with the restricted Coulomb graph, on which $y_i^2=0$.
The result is thus to be contrasted with the massless limit of the octagon, in which the \emph{internal} masses are set to zero first and in which one finds a different kind of exponentiation \cite{Coronado:2018cxj,Belitsky:2019fan}\footnote{The one-loop exact term is only valid at weak coupling, while at strong coupling it gets modified to become linear in $g$.}:
\bba
\lim_{x_{i,i+1}^2, d_{13},d_{24}\to 0}\log M &= \lim_{x_{i,i+1}^2, d_{13},d_{24}\to 0} \log\mathbb{O}\nonumber\\
&=-\frac{\Gamma_{\rm oct}(g)}{16}\,\log^2\left(\frac{x_{12}^2 x_{23}^2 x_{34}^2 x_{41}^2}{\left(x_{13}^2 x_{24}^2\right)^2}\right)-\frac{g^2}{4}\,\log^2\left(\frac{x_{12}^2 x_{34}^2}{x_{23}^2 x_{41}^2}\right) -\frac{1}{2}D_{0}(g)\,,
\end{align}
where
 \bba
 \Gamma_{\rm oct}(g) &= \frac{2}{\pi^2}\,\log\cosh(2\pi g)\overset{g\to 0}{=} 4\,g^2 -16\zeta_{2}\,g^4+256\zeta_4\,g^6-3264\zeta_{6}\,g^8 +\OOk(g^{10})\,.
 \end{align}
We see that not all massless limits of scattering amplitudes are controlled by $\Gamma_{\rm cusp}$!
It would be interesting to understand how to interpolate between these approaches to the massless limit.

\section{A Coulomb branch amplitude from integrability}
\label{sec:Integrability}

An alternative to the hard problem of evaluating high-loop order conformal integrals is to use the hexagonalization of \cite{Fleury:2016ykk}  to find the four-point functions analytically. This is an integrability-based method which provides  a finite-coupling prescription to compute correlation functions. In this section we use this method to provide a finite-coupling result for the generating function of simplest correlators which, through the duality \eqref{eq:O=M}, also gives the coulomb branch amplitude. Our result is given in terms of the octagon form factors $\mathbb{O}_{l}$, introduced in \cite{Coronado:2018ypq,Coronado:2018cxj} and vastly studied at weak, strong and finite coupling in \cite{Kostov:2019stn,Kostov:2019auq,Belitsky:2019fan,Bargheer:2019kxb,Bargheer:2019exp,Belitsky:2020qrm,Belitsky:2020qir,Belitsky:2020qzm,Kostov:2021omc}. Furthermore, we consider the weak coupling limit of the octagons and compare it with the integrand of the coulomb branch amplitude, which can be found up to ten loops thanks to the  $10D$ uplift \eqref{inverse} of the four-gluon integrand given in \cite{Bourjaily:2016evz}. This allows us to find a map between simple analytic functions and linear combinations of conformal integrals, most of them unknown at the function level.

\subsection{Hexagonalization and octagons}

\begin{figure}[t]
 \begin{minipage}[h]{1\textwidth}
 \centering 
  \resizebox{3\totalheight}{!}{\includestandalone[width=1\textwidth]{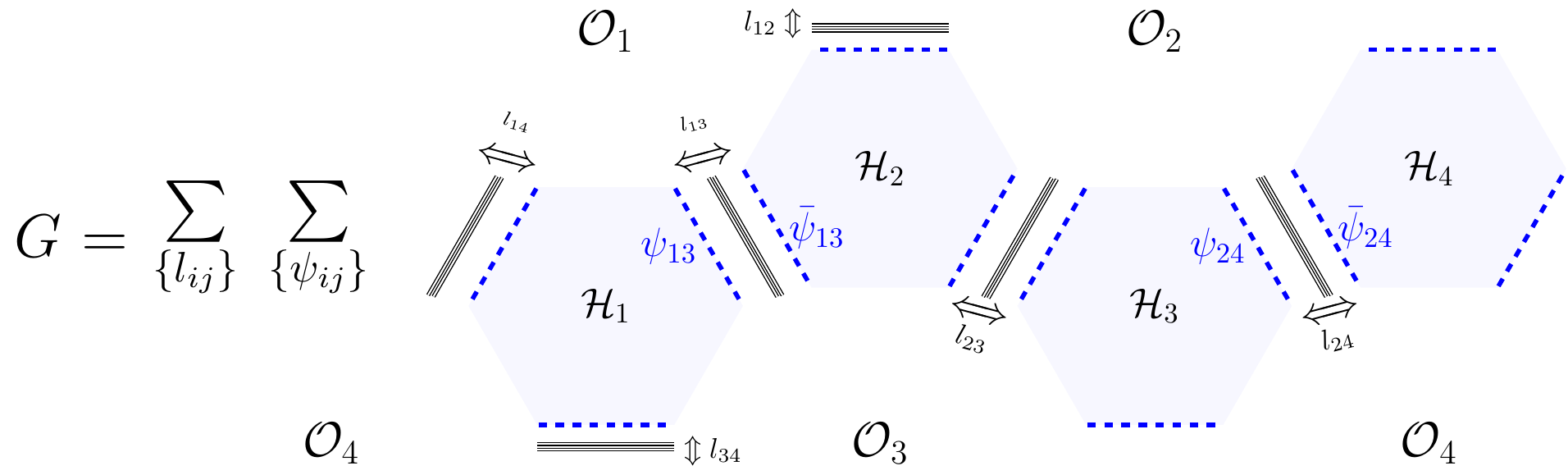}}
 \end{minipage}
 \caption{Hexagonalization: finite-coupling correlators are obtained from the sum over the free theory graphs (skeleton graphs) labeled by the six bridges  $\{l_{ij}\}$, tessellated into four hexagons $\mathcal{H}_{\psi_{ij},\psi_{jk},\psi_{ki}}$ which are glued by sums over complete basis of mirror states $\{\psi_{ij}\}$ that propagate traversing the bridges. The unrestricted sum over all non-negative integers $l_{ij}$ give the sum $G$ of all four-point planar correlators.}
 \label{fig:Hexagonalization}
 \end{figure}

Hexagonalization \cite{Eden:2016xvg,Fleury:2016ykk} proposes an uplift from the free theory to the finite-coupling correlator by dressing the free theory graphs with hexagons \cite{Basso:2015zoa} as shown in fig.~\ref{fig:Hexagonalization}.  According to ref.~\cite{Fleury:2016ykk}:
\beq\label{eq:Hexagonalization}
G  = \sum_{\{l_{ij}\}} \underbrace{\,c_{\{l_{ij}\}}\,\prod_{(i,j)}(d_{ij})^{l_{ij}}}_{\text{free Wick contractions}}\left[\sum_{\{\psi_{ij}\}}\prod_{(i,j)}\mu_{\psi_{ij}}\,e^{-E_{\psi_{ij}}l_{ij}}\,\mathcal{W}_{\psi_{ij}}\,\prod_{(i,j,k)}\,\mathcal{H}_{\psi_{ij},\psi_{jk},\psi_{ki}}\right]
\eeq
where we consider a sum over all positive bridges $l_{ij}>0$ to account for all possible charges and obtain the generating function $G$ of all four-point functions. 

\begin{figure}[t]
 \begin{minipage}[h]{1\textwidth}
 \centering 
  \resizebox{4\totalheight}{!}{\includestandalone[width=1\textwidth]{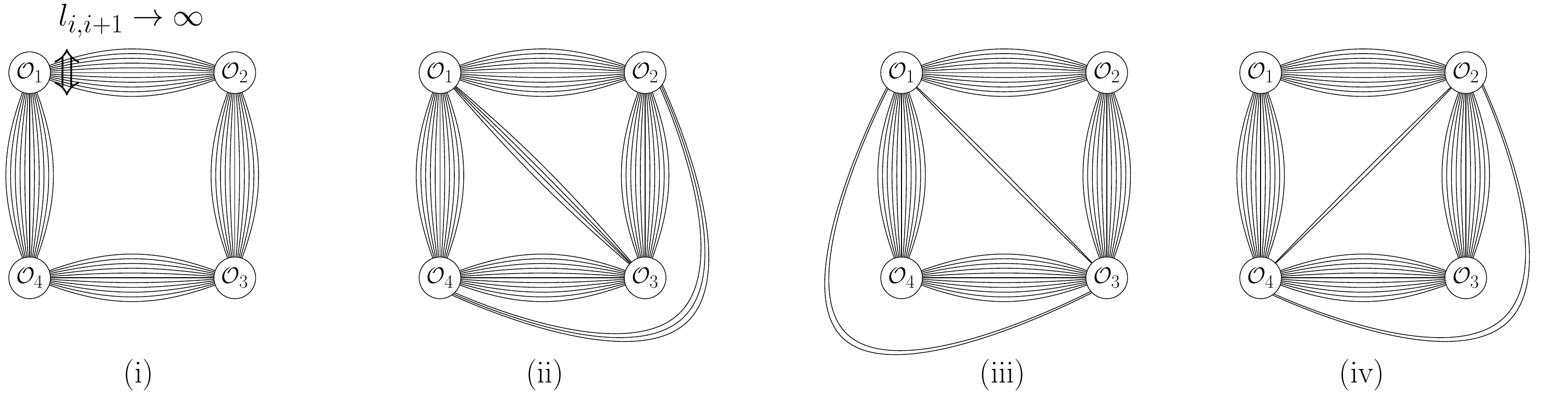}}
 \end{minipage}
 \caption{The four types of skeleton graphs contributing to the large R-charge correlators or 10D null limit. All graphs have a square perimeter with bridges of infinite length and they are distinguished by  the position and length of their diagonal bridges.}
 \label{fig:SimpleSkeletons}
 \end{figure}

In principle, all the ingredients in \eqref{eq:Hexagonalization} are known at finite coupling, allowing to perform the infinite sum and obtain $G$. However, in practice this is a complicated task which first requires dealing with the involved matrix structure of the hexagon form factors $\mathcal{H}_{\psi_{ij},\psi_{jk},\psi_{ki}}$  for generic mirror states $\psi_{ij}$\footnote{see for instance \cite{Jiang:2016ulr,Basso:2019diw} for resummations at strong coupling in the context of three-point functions  and see \cite{Fleury:2017eph,deLeeuw:2019qvz,Fleury:2020ykw} for examples on more generic hexagon form factors in the context of four- and five-point functions at weak coupling.}. Fortunately this procedure simplifies when taking the 10D null limit since this only receives contributions from skeleton graphs with large R-charges in their perimeter, see fig.~\ref{fig:SimpleSkeletons}. For these graphs the sum over states propagating on the bridges with infinite length $l_{12},l_{23},l_{34},l_{41}$ projects to the vacuum state thanks to the exponentially suppressed factor in \eqref{eq:Hexagonalization}. This results on a factorization into two building blocks that we call octagons.

The octagon form factor is composed by a pair of hexagons separated by a diagonal bridge $l$ and glued by a sum over a complete basis of mirror states propagating on this bridge. Schematically we represent it as:
\beq
\text{Octagon}_{l}(z,\bar{z},\alpha,\bar{\alpha})\,=\,\sum_{{\color{blue}\psi}}\,\raisebox{-2077787sp}{\resizebox{.3\totalheight}{!}{\includestandalone[width=.6\textwidth]{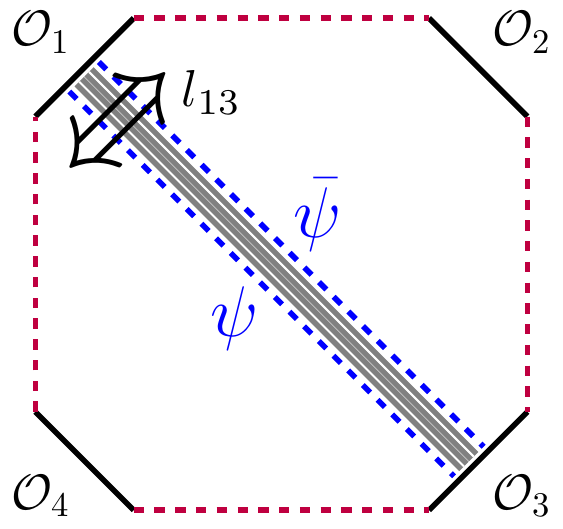}}}\,
\eeq
Explicit finite-coupling expressions for this octagon form factor can be found in \cite{Coronado:2018ypq} in terms of the spacetime $(z,\bar{z})$ and R-charge $(\alpha,\bar{\alpha})$ cross-ratios defined as in \eqref{eq:crossratios}.  We review some of these representations in section \ref{sec:Ofinite}. We are interested in the 10D null limit obtained by simply  setting $y_{i,i+1}^2=-x_{i,i+1}^2$:
\bba
\mathbb{O}_{l} &= \lim_{d_{i,i+1}\to 1}\text{Octagon}_{l}(z,\bar{z},\alpha,\bar{\alpha})
\nonumber\\
&=\text{Octagon}_{l}(z,\bar{z},\alpha_*,\bar{\alpha}_*)\,,
\end{align}
where the R-charge cross-ratios solve the equations:
$\frac{\alpha_*\bar{\alpha}_*}{z\bar{z}} =
\frac{(1-\alpha_*)(1-\bar{\alpha}_*)}{(1-z)(1-\bar{z})}=\frac{1}{d_{13}d_{24}}$.
The object $\mathbb{O}_l$ serves as a building block to compute our 10D null octagon as the sum:
 \beq\label{eq:sumObridge}
\boxed{ \mathbb{O} = \mathbb{O}_{0} + \sum_{l=1}^{\infty}(d_{13})^{l}\,\mathbb{O}_{l} +(d_{24})^{l}\,\mathbb{O}_{l}\,. }
\eeq
This result has the structure dictated by hexagonalization: each tree level graph, represented by the diagonal propagators, gets dressed by the building block $\mathbb{O}_{l}$.

It is instructive to verify that squaring this expression reproduces
the generating function of all simplest correlators,
 which can be organized following the rules of hexagonalization as:
\bba\label{eq:sumoctagons2}
\mathbb{O}\times\mathbb{O} & =\sum_{l_{13},l_{24}=0}^{\infty}\,(d_{13})^{l_{13}}\, (d_{24})^{l_{24}} \, \bigg{(}\delta_{l_{13},0}\delta_{l_{24},0}\,\mathbb{O}_{0}\,\mathbb{O}_{0} +2(1-\delta_{l_{13},0})(1-\delta_{l_{24},0})\,\mathbb{O}_{l_{13}}\mathbb{O}_{l_{24}} \nonumber\\
&\qquad\quad+(1-\delta_{l_{13},0})\delta_{l_{24},0}\sum_{m=0}^{l_{13}}\mathbb{O}_{m}\mathbb{O}_{l_{13}-m} +\delta_{l_{13},0}(1-\delta_{l_{24},0})\sum_{m=0}^{l_{24}}\mathbb{O}_{m}\mathbb{O}_{l_{24}-m}\bigg{)}\,.
\end{align}
This is an example of ``octogonalization": the four terms in the sum \eqref{eq:sumoctagons2} are associated to the four types of skeleton graphs, listed in fig.~\ref{fig:SimpleSkeletons}, dressed by a pair of octagons $\mathbb{O}_{l}$.  The first term corresponds to the case  without diagonal bridges $l_{13}=l_{24}=0$ in fig.~\ref{fig:SimpleSkeletons} ({\romannumeral 1}). The second term has both non-zero diagonals $l_{13},l_{24} \neq 0$, hence the accompanying factor of 2 corresponding to the swapping of the bridges positions between the inside and the outside of the square frame in the graph ({\romannumeral 2}). Finally, the third and fourth terms  correspond to the graphs  ({\romannumeral 3}) and  ({\romannumeral 4}) which have a single non-zero diagonal bridge, $l_{13}\neq0, l_{24}=0$ and $l_{13}=0, l_{24}\neq 0$ respectively. This condition allows the single diagonal bundle to split between the inside and outside of the square frame while still respecting planarity and explains the  sums  in the bottom line of \eqref{eq:sumoctagons2}.

Thanks to the duality \eqref{eq:O=M}, our integrability-based result for the 10D null octagon \eqref{eq:sumObridge} provides a finite coupling representation of a finite Coulomb branch amplitude in terms of $\mathbb{O}_{l}$. This allows for a thorough study of the amplitude in various regimes of the coupling and kinematics using the results in \cite{Coronado:2018ypq,Coronado:2018cxj,Kostov:2019stn,Kostov:2019auq,Belitsky:2019fan,Bargheer:2019kxb,Bargheer:2019exp,Belitsky:2020qrm,Belitsky:2020qir,Belitsky:2020qzm,Kostov:2021omc}. In this paper we follow a different route and take advantage of the knowledge of the amplitude integrand $M$ up to 10 loops and compare it with the results from integrability in the weak coupling limit.  For this purpose, in what follows we review the representations of the octagon form factor and its weak coupling limit.

\subsection{The octagon from integrability: 10D null limit and weak coupling expansion}
\label{sec:FiniteOctagon}

In \cite{Coronado:2018ypq}, using integrability, the octagon form factor with bridge $l$ was found as a sum over mirror states:
\beq
\text{Octagon}_{l}(z,\bar{z},\alpha,\bar{\alpha}) = 1+ \sum_{n=1}^{\infty}\chi_{n}(z,\bar{z},\alpha,\bar{\alpha})\,I_{n,l}(z,\bar{z})
\eeq
where $n$ stands for the number of mirror particles. This $n$-particle contributions can be decomposed into a rational factor $\chi_{n}$ carrying all the $R$-charge dependence and the $n$-particle integrals $I_{n,l}$ depending only on the spacetime cross ratios. Explicit representations for these components, at finite coupling, can be found in \cite{Coronado:2018ypq}. Here it only suffices to know the 10D null limit:
\beq
\lim_{d_{i,i+1}\to 1} \chi_{n}\,=\, (1-d_{13}d_{24})^{n}
\eeq
to obtain: 
\beq\label{eq:O10Dnull}
\boxed{ \mathbb{O}_{l}(z,\bar{z},d_{13},d_{24})= 1+\sum_{n=1}^{\infty}(1-d_{13}d_{24})^{n}\,I_{n,l}(z,\bar{z})\,. }
\eeq
An alternative and more efficient way to calculate the octagon is through the determinant of an infinite matrix \cite{Kostov:2019stn,Kostov:2019auq,Belitsky:2019fan,Belitsky:2020qrm,Belitsky:2020qir,Belitsky:2020qzm,Kostov:2021omc}:
\beq
\mathbb{O}_{l} \,=\, \det(1-\mathbb{K}_{l})
\eeq
whose elements are given by integrals of Bessel functions of the first kind:
\beq\label{eq:Kmat}
\left(\mathbb{K}_{l}\right)_{ij}\,=\,2(-1)^{i-j}(2j+l-1)\,\int_{0}^{\infty}d\tau\,\chi(\tau)\,\frac{J_{2i+l-1}(2 g\tau)\,J_{2j+l-1}(2g\tau)}{\tau}
\eeq
where $i,j\geq 1$ and only the weight factor $\chi(\tau)$ depends on the spacetime and R-charge cross ratios. In our 10D null limit this weight factor becomes effectively:
\bba
\chi(\tau)&=\frac{(1-d_{13}d_{24})}{\sqrt{z \bar{z}(1{-}z)(1{-}\bar{z})}}\frac{1}{\cosh(\sqrt{\zeta^2+\tau^2})-\cos\phi}\,,\text{ with }e^{-2\zeta}=\frac{z\bar{z}}{(1{-}z)(1{-}\bar{z})}\,,\,e^{2i\phi}=\frac{z(1{-}\bar{z})}{\bar{z}(1{-}z)}\,.
\end{align}
The integrals in \eqref{eq:Kmat} can be easily evaluated at weak coupling as we now review.

\subsubsection{Octagon at weak coupling}
\label{sec:Ofinite}

In this regime, the infinite matrix $\mathbb{K}_{l}$ can be truncated to a matrix of dimension $\frac{\ell-l}{2}$ when interested in the $\ell$-loop order. This is because the Bessel functions scale as $J_{n}(g\tau)\propto (g\tau)^{n}$. Furthermore, the integral over $\tau$ can be easily perform order by order in the coupling: the integration of $\chi(\tau)\tau^{2n-1}$ is proportional to the well-known ladder integral $ F_{n}$. These functions were introduced in \cite{Usyukina:1993ch} as:
\beq\label{eq:Fladder}
F_{p}(z,\bar{z}) =
(-1)\sum_{j=p}^{2p}\frac{j!\left[-\log(\frac{z}{z-1}\frac{\bar{z}}{\bar{z}-1})\right]^{2p-j}}{p!(j-p)!(2p-j)!}\left[\frac{\text{Li}_{j}(\frac{z}{z-1})-\text{Li}_{j}(\frac{\bar{z}}{\bar{z}-1})}{z-\bar{z}}\right].
\eeq
In this basis, the expansion of the matrix elements goes as:
\beq
\left(\mathbb{K}_{l}\right)_{ij} = \sum_{\ell =i+j+l-1}^{\infty}\,C_{l;i,j}^{(\ell)}\,(-g^{2})^{\ell}\,(1-d_{13}d_{24})\,F_{\ell}
\eeq
with coefficients:
\beq
C_{l;i,j}^{(\ell)}\,=\, \frac{(-1)^{l+1}(2j+l-1)(2\ell)!\,\ell!\,(\ell-1)!}{(\ell-(i+j+l-1))!\,(\ell+(i+j+l-1))!\,(\ell-|i-j|)!\,(\ell+|i-j|)!}\,.
\eeq
The coefficient of $(1-d_{13}d_{24})^{n}$ can thus be loop-expanded as:
\beq\label{eq:nparticleWeak}
I_{n,l}=\sum_{\ell=n(n+l)}^{\infty}(-g^{2})^{\ell}\,I^{(\ell)}_{n,l}
\eeq
where each loop correction is a sum of products of $n$ ladder integrals.
The basis of such products can be further refined: they only appear in combinations such that the double
discontinuity around $z\to\infty$ vanishes.\footnote{This property can be seen directly from eq.~\eqref{eq:Kmat} because monodromy around $z\to\infty$ shifts $\mathbb{K}_{ij}$ by the residue of a pole at the value
$t_*$ where the denominator of $\chi(t)$. vanishes. This residue is a matrix of rank one.}
This forces the ladders to appear in determinants which generalize those in the so-called fishnet diagrams \cite{Basso:2017jwq}.  A basis of such determinants was introduced in \cite{Coronado:2018cxj} as:
\beq\label{eq:Mmatrices}
F_{\color{blue}i_{1},i_{2},\cdots, i_{n}} =\prod_{m=1}^{n}\frac{1}{i_{m}!(i_{m}-1)!}  \begin{vmatrix}
\color{blue} f_{i_{1}} & f_{i_{2}-1}  & \cdots  & f_{i_{n}-n+1} \\
f_{i_{1}+1} & \color{blue} f_{i_{2}} & \cdots  & f_{i_{n}-n+2}\\
\vdots & \vdots & \ddots & \vdots\\
f_{i_{1}+n-1} & f_{i_{2}+n-2} & \dots & \color{blue} f_{i_{n}}\\
\end{vmatrix}\qquad\text{with }f_{p} \,=\,p!(p-1)!\, F_{p}
\eeq 
and the diagonal elements are dictated by the subscripts in $F_{i_{1},i_{2},\cdots,i_{n}}$. Under this definition, the elements of our basis with a single index coincide with the standard ladder integral \eqref{eq:Fladder}.

Using this basis, the $\ell$-loop contribution can be written as
a sum of determinants with $\ell$ indices and strong restrictions:
\beq\label{eq:SumSteinmann}
I^{(\ell)}_{n,l}= (-1)^{n\,l}\times\sum_{i_{1}+\cdots + i_{n}=\ell \atop \text{with }(i_{p+1}-i_{p})\geq 2\text{ and }i_{1}> l} c_{l\,;\{i\}_{n}} \,F_{i_{1},i_{2},\cdots,i_{n}}\,.
\eeq
The coefficients  $c_{l\,;\{i\}_{n}}$ are positive rational numbers for which we have not found a closed form expression, except for the  one-particle case:
\beq\label{eq:oneparticle}
I^{(\ell)}_{n=1,l} = (-1)^{l}\,\binom{2\ell-2}{\ell-l-1}\,F_{\ell}\,.
\eeq
In \cite{Coronado:2018cxj} it was shown how to fix all coefficients $c_{l\,;\{i\}_{n}}$ for arbitrary $n$ and bridge $l=0$ following a bootstrap approach that uses special kinematic limits of $\mathbb{O}_{0}$. Unfortunately, we lack a bootstrap method to fix the coefficients with $l>0$. For these we resort to a direct loop expansion of the integrability results. 

In  table \ref{tab:twoparticleDets} we show all non-vanishing loop corrections below 9 loops. Furthermore, by performing expansions up to 25 loops ($n(n+l)\leq 25$), we have identified that the leading and sub-leading loop contributions  for arbitrary $n$ and $l$ are each given by a single determinant. The leading term is:
\beq\label{eq:leadingS}
I^{[n(n+{\color{blue}l})]}_{n,{\color{blue}l}} =(-1)^{n\,l}\,F_{1+{\color{blue}l},\,3+{\color{blue}l},\,\cdots,2n-1+{\color{blue}l}}
\eeq
and the subleading term carries the same indexes except for the last one which is shifted by one unit:
\beq\label{eq:subleadingS}
I^{[n(n+{\color{blue}l})+{\color{red}1}]}_{n,{\color{blue}l}} = (-1)^{n\,l}\,2(2n-1+l)\,F_{1+{\color{blue}l},\,3+{\color{blue}l},\,\cdots,\,2n-1+{\color{blue}l}+{\color{red}1}}\,.
\eeq
These leading and subleading corrections will be mapped to scattering amplitude integrals shortly.


\begin{table}
\begin{minipage}[t]{.5\textwidth}
 \centering 
\def\arraystretch{1.1}
\resizebox{2.5\totalheight}{!}{
\begin{tabular}{|c|cccccc|ccc|cc|}
\hline
\diagbox[height=1.5\line]{\raisebox{-3ex}{\Large $\overset{\ell}{\,}$}}{\raisebox{3ex}{\Large $\underset{l}{\,}$}} & & & $\mathbb{O}_{0}$ & &  & & & \qquad \qquad $\mathbb{O}_{1}$ &   & \qquad $\mathbb{O}_{2}$ &     \\   \hline
1 & $ F_1$ &                                          &                                         &                                              &                                            & &  0&             &                                                                          & 0 &                               \\   \hline
2 & $ 2 F_2$ &                                          &                                         &                                              &                                             & & $-F_{2}$     &                                             &                           & 0  &                             \\   \hline
3 & $ 6F_3$ &                                          &                                         &                                              &                                              & &$-4F_{3}$     &                                             &                              & $F_3$ &                       \\    \hline
4 & $20F_4$ & $ \color{blue}+F_{1,3}$     &                                         &                                              &                                               & & $-15F_{4}$   &                                              &                             & $6F_4$  &                                \\   \hline
5 & $70F_5$ & $ \color{blue} +6\,F_{1,4}$ &                                         &                                              &                                              & &  $-56F_{5}$   &                                              &                           & $28F_5$ &                                    \\   \hline
 6 & $252F_6$ &  $+28\,F_{1,5}$                &$+\frac{18}{5}\,F_{2,4}$    &                                              &                                              & &  $-210F_{6}$   &   $ \color{blue}+F_{2,4}$        &                            & $120F_6$ &                                       \\    \hline
 7 & $924F_7$ & $+120\,F_{1,6}$                &$+24\,F_{2,5}$                   &                                               &                                             & &  $-792F_{7}$  &   $ \color{blue}+8\,F_{2,5}$   &                                 &  $495 F_7$ &                                                \\    \hline
8 & $3432 F_8$ &  $+495\,F_{1,7}$               &$+\frac{855}{7}\,F_{2,6} $ & $+\frac{162}{7} F_{3,5}$       &                                             & &   $-3003F_{8}$ &    $+45\,F_{2,6}$                 & $+11\,F_{3,5}$                &    $2002 F_8$ & $ \color{blue}+F_{3,5}$   \\    \hline
9 &  $12870 F_9$ & $+2002\,F_{1,8}$             & $+561\,F_{2,7} $                & $ +\frac{2325}{14}\,F_{3,6}$ & ${\color{blue}+F_{1,3,5}}$  & &  $-11440 F_{9}$ & $+220\,F_{2,7}$                     & $+\frac{600}{7}\,F_{3,6}$ &  $8008F_9$&  $ \color{blue}+10\,F_{3,6}$    \\ \hline
\end{tabular}
}
 \end{minipage}
 \caption{Octagons up to nine loops for bridges $l=0,1,2$. In order to restore the coupling and  $10D$ character  we make the rescaling $F_{i_1,\cdots, i_{n}}\to(-g^2)^{i_1+\cdots + i_{n}}(1-d_{13}d_{24})^{n}\,F_{i_1,\cdots, i_{n}}$. We highlight the single-determinant cases given by the close formulas
\eqref{eq:leadingS}-\eqref{eq:subleadingS}.}
 \label{tab:twoparticleDets}
 \end{table}

\subsection{Integrals from octagons}
\label{sec:comparison}

In this section we compare the representation of the coulomb branch amplitude given by conformal integrals against the integrability representation of the octagon.
The integrand can be obtained in either of two ways: by taking the 10D null limit of the correlation
function integrand from eq.~\eqref{inverse7}, or alternatively, by uplifting to $10$-dimensions
directly the known integrand for massless gluon amplitudes using the same procedure:
\beq
M(x,y)\Big|_{\rm integrand} \, = \, M(x,0)\big{|}_{x_{13}^2\,\mapsto X_{13}^2,\ x_{24}^2\,\mapsto X_{24}^2}\,.
\eeq
As noted, this uplift is unambiguous up to and including at least seven loops, where Gram determinants ambiguities
were shown above to be absent.

Integrated expressions for $M$ on the other hand follow directly
from our new conjectured amplitude/correlator duality  \eqref{eq:O=M} and the computation of the octagon $\mathbb{O}$ using integrability:
\bba\label{eq:M=sumIn}
M= \frac{\mathbb{O}}{\mathbb{O}^{\text{free}}}=
1+ (1-d_{13})(1-d_{24})\sum_{l=0}^{\infty}
\frac{(d_{13})^l+(d_{24})^l}{1+\delta_{l,0}}\sum_{n=1}^{\infty}(1-d_{13}d_{24})^{n-1}\,I_{n,l} 
\end{align}
where we made use of \eqref{eq:O10Dnull} and
 the free limit: $\mathbb{O}^{\text{free}} = 1+ \sum\limits_{l=1}^{\infty}(d_{13})^l+(d_{24})^l = \frac{1-d_{13} d_{24}}{(1-d_{13})(1-d_{24})}$.

To compare these two formulas, it is helpful to group terms with different dependence on
$d_{13}$ and $d_{24}$:
\beq\label{eq:Mab}
M
=1+ \sum_{a,b\geq 1} (1-d_{13})^a\,(1-d_{24})^b\,M_{a,b}\,.
\eeq
Using the formulas recorded in \eqref{eq:M3}, we can readily check the agreement up to three loops:
\bba
M_{1,1} &=-g^{2}\,x_{13}^2x_{24}^2\,g_{1234}  &=& - g^2\, F_1\,, \label{M11 exact} \\
M_{1,2} &=g^{4}\, x_{13}^2x_{24}^2\,h_{13;24} \,-\, g^{6}\,x_{13}^2x_{24}^2\,(T_{13;24}+T_{13;42})\,+\,\mathcal{O}(g^8) &=&\, g^{4}\,F_{2}-2\,g^{6}\,F_{3}+\,\mathcal{O}(g^8)\,, \nonumber \\
M_{2,1} &=g^4 x_{13}^2x_{24}^2\,h_{24;13} - g^6 x_{13}^2x_{24}^2\,(T_{24;13}+T_{24;31})\,+\,\mathcal{O}(g^8)&=&\, g^{4}\,F_{2}-2\,g^{6}\,F_{3}+\,\mathcal{O}(g^8)\,, \nonumber \\
 M_{1,3} &= -g^{6}\,x_{13}^2x_{24}^2\,L_{13;24}+\,\mathcal{O}(g^8) &=&\,-g^{6}\,F_{3}+\,\mathcal{O}(g^8)\,,\nonumber \\
 M_{3,1} &= -g^{6}\,x_{13}^2x_{24}^2\,L_{24;13} +\,\mathcal{O}(g^8) &=&-g^{6}\,F_{3}+\,\mathcal{O}(g^8)\,.
\end{align}
All integrated expressions satisfy the symmetry $M_{a,b}=M_{b,a}$, which is a simple
consequence of conformal invariance since all cross-ratios are invariant \cite{Drummond:2006rz}.
Note that $M_{1,1}$ is one-loop exact: this is easy to understand from the scattering amplitude picture (all conformal integrals with $\ell\geq 2$ loops require more powers of $x_{13}^2$ or $x_{24}^2$ in the numerators).

In general, by equating \eqref{eq:M=sumIn} with \eqref{eq:Mab}
we find the following expression  for the partial amplitudes $M_{a,b}$ in terms of the
$n$-particle integrability contributions $I_{n,l}$:
\beq\label{eq:Map}
M_{a,b} \,=\,\sum_{n=\min(a,b)}^{a+b-1}\,\sum_{\substack{l\geq0\\ l+n\geq a,b}}^{\infty}  \,C^{n,l}_{a,b}\,I_{n,l}
\eeq
with
\beq
C^{n,l}_{a,b} =\frac{(-1)^{a+b-n-1}}{1+\delta_{l,0}}\,\left[\binom{n-1}{a-1}\,\binom{a+l-1}{a+b-n-1} \,+\,(a\leftrightarrow b)\right].
\eeq
While eq.~\eqref{eq:Map} is valid at finite coupling, here we will only study it order by order in the coupling $g^2$.  We have seen in eq.~\eqref{eq:SumSteinmann}
that the right-hand-side, to all orders in the coupling, is a sum of determinants of ladder integrals.
This is far from obvious from the integrals on the left-hand-side and this predicts new integral relations.

The simplest case corresponds to $M_{1,1}$ where only $n=1$ particles contribute in eq.~\eqref{eq:Map}:
\beq
M_{1,1} =  I_{1,0}+2\sum_{l=1}^{\infty} I_{1,l}\,.
\eeq
The infinite sum from integrability agrees nontrivially with
the one-loop exact result $M_{1,1} = g^{2} f_{1}$ quoted in eq.~\eqref{M11 exact}.

Groups  $M_{a,b}$ with $b\geq 2$ receive contributions from two-particle integrals, starting from high-enough loop order. For instance for $a=1$ we have:
\beq
M_{1,b} \,=(-1)^{b-1}\underbrace{\left[(-g^{2})^{b}\,I_{1,b-1}^{(b)}+(-g^{2})^{b+1}\left(I_{1,b-1}^{(b+1)}+b\,I_{1,b}^{(b+1)}\right)+\cdots\right]}_{\text{single-particle}}+(-g^{2})^{2b}\,I_{2,b-2}^{(2b)}+\mathcal{O}(g^{4b+2}) 
\eeq
where the upper parenthesis $I^{(\ell)}$ denote the coefficient of $(-g^{2})^{\ell}$.
From the viewpoint of the scattering amplitudes, the leading term is simply a single ladder integral:
\beq
  M_{1,b}^{(b)} \,=\, \raisebox{-1577787sp}{\resizebox{3\totalheight}{!}{\includestandalone[width=.6\textwidth]{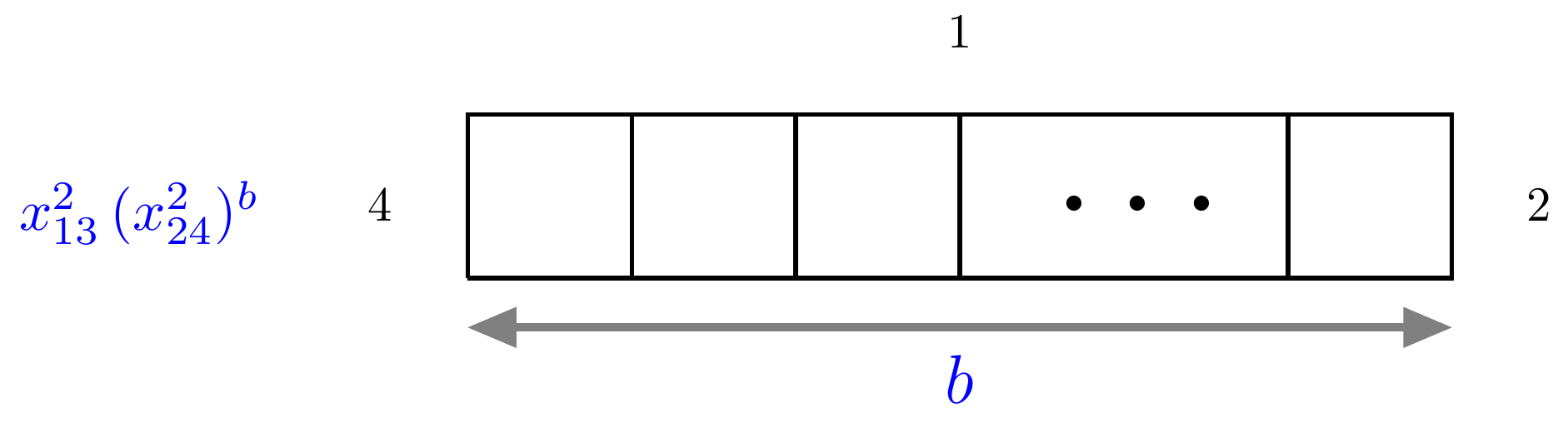}}}=\,F_b\,.
\eeq
The ladder integrals were famously computed long ago \cite{Usyukina:1993ch} and one might say that
our duality provides one (more) re-derivation through integrability.

The single-particle contributions can be projected out by considering partial amplitudes with $a,b\geq 2$.
The leading contribution sets in at order $M_{a,b}\sim g^{2ab}$ where the integrand
is simply the fishnet integral of ref.~\cite{Basso:2017jwq,Basso:2021omx} which will be discussed below.
Both the leading and subleading term are given by a single term with $n=a$ and $l=b-a$ in the integrability formula:
\beq\label{eq:FishandSubFish}
M_{a,b} \,=\, (-g^{2})^{a b}\,I^{(a b)}_{a,b-a}+(-g^{2})^{a b+1}\,I^{(a b+1)}_{a,b-a}+\mathcal{O}(g^{2(ab+2)})\;\;\text{with }b\geq a\geq 2\,.
\eeq
Interestingly, each of these evaluates to a single determinant of ladder integrals, see eq.~\eqref{eq:subleadingS}.
The contributing amplitude integrals are nontrivial and will be discussed below.

\subsubsection{Detailed comparison at four-loops}\label{app:fourloop}

\begin{figure}[t]
\centering
 \resizebox{1.05\totalheight}{!}{\includestandalone[width=1.5\textwidth]{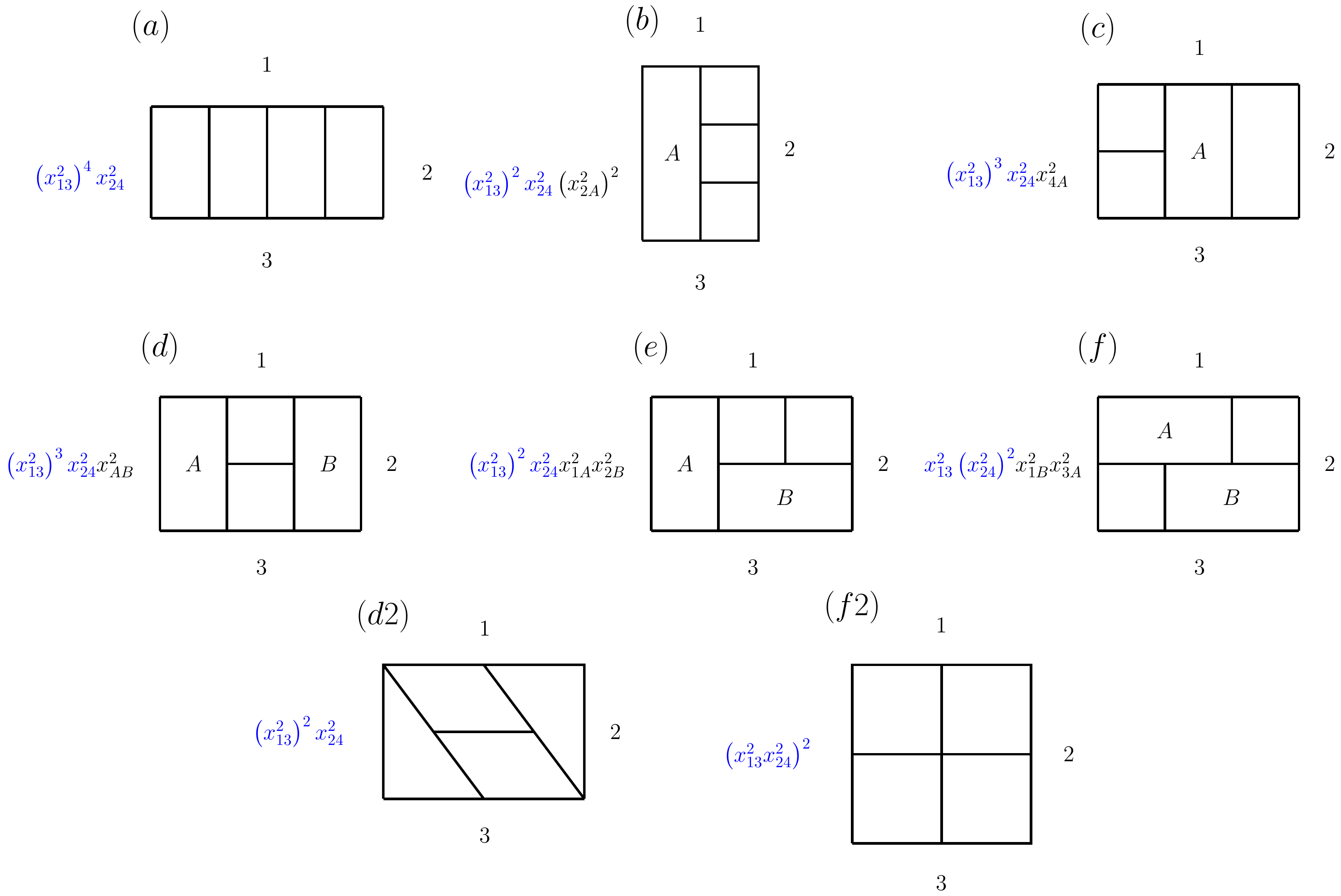}}
 \caption{Four-loop integrals entering eq.~\eqref{eq:M22} in scattering amplitude notation.}
\label{fig:FourLoopAmplitude}
\end{figure}

The comparison between Feynman integrals and integrability becomes particularly interesting starting from four-loops: non-ladder integrals appear at the same time as multi-particle contributions.

Evaluating the amplitude \eqref{eq:Map} using the integrability results quoted in eqs.~\eqref{eq:oneparticle} and \eqref{eq:leadingS} we find the following predictions:
\bba\label{eq:Map4loop}
\qquad\qquad\qquad\qquad M_{4,1}^{(4)}  &= -I_{1,3}^{(4)}  &=&\,F_4\,, \nonumber\\
M_{3,1}^{(4)} &= I_{1,2}^{(4)} +3 I_{1,3}^{(4)} &=&\, 3\,F_4\,, \nonumber\\
M_{2,1}^{(4)}  &=-I_{1,1}^{(4)} -2I_{1,2}^{(4)} -3I_{1,3}^{(4)}+ I_{2,0}^{(4)} &=&\,6\,F_{4}\,+\,F_{1,3}\,,\nonumber\\
M_{2,2}^{(4)} &= -I_{2,0}^{(4)} &=&\,  -F_{1,3}\,,\qquad\qquad\qquad\qquad
\end{align}
where $F_{1,3}=\frac{1}{12}\begin{vmatrix} f_{1} & f_{2} \\ f_{2} & f_{3} \end{vmatrix}$
is the simplest determinant of ladders.

The same amplitudes are also expressed as the following Feynman integrals \cite{Bern:2006ew},
shown in fig.~\ref{fig:FourLoopAmplitude}:
\bba
M^{(4)}_{4,1}&=\mathcal{I}^{(4)a}_{1,2,3,4}\,,\nonumber\\
M^{(4)}_{3,1}&= \left(\mathcal{I}^{(4)c}_{1,2,3,4}+ \mathcal{I}^{(4)c}_{1,4,3,2}\right)+ \mathcal{I}^{(4)d}_{1,2,3,4}\,,\nonumber\\
M^{(4)}_{2,1}&=\left( \mathcal{I}^{(4)b}_{1,2,3,4}+ \mathcal{I}^{(4)b}_{1,4,3,2}\right)+ \left(\mathcal{I}^{(4)e}_{1,2,3,4}+ \mathcal{I}^{(4)e}_{1,4,3,2}+ \mathcal{I}^{(4)e}_{3,2,1,4}+\mathcal{I}^{(4)e}_{3,4,1,2}\right)-\left(\mathcal{I}^{(4)d2}_{1,2,3,4}+\mathcal{I}^{(4)d2}_{1,4,3,2}\right)\,,\nonumber\\
&\qquad +\left(\mathcal{I}^{(4)f}_{2,1,4,3}+\mathcal{I}^{(4)f}_{2,3,4,1}\right)\,,\nonumber\\
M^{(4)}_{2,2}&= \,-\,\mathcal{I}^{(4)f2}_{1,2,3,4}\,.
\label{eq:M22}
\end{align}
The parentheses group the permutations of each integral.
Note that here we use the scattering amplitude notation; corresponding conformal
integrals follow simply from the graph duality illustrated in fig.~\ref{fig:3loopAmplitude}.

We now equate the two representations \eqref{eq:Map4loop} and \eqref{eq:M22}.
Since the integrals are evaluated off-shell $(x_{i,i+1}^2\neq 0)$ and are finite, we can use
 conformal symmetry and magic identities to relate the first five conformal integrals in fig.~\ref{fig:FourLoopAmplitude}, with any order of their indexes, to a unique ladder integral:
\beq
\mathcal{I}^{(4)a} =\mathcal{I}^{(4)b} =\mathcal{I}^{(4)c} =\mathcal{I}^{(4)d} =\mathcal{I}^{(4)e} = F_{4} \,.
\eeq
Thanks to these identities, the first two lines of eq.~\eqref{eq:M22} trivially reproduce the first two of eq.~\eqref{eq:Map4loop}.  The remaining comparisons demand the following two integral identities:
\bba\label{eq:NewIdentity}
-\left(\mathcal{I}^{(4)d2}_{1,2,3,4}+\mathcal{I}^{(4)d2}_{1,4,3,2}\right)\,+\,\left(\mathcal{I}^{(4)f}_{2,1,4,3}+\mathcal{I}^{(4)f}_{2,3,4,1}\right) = \mathcal{I}^{(4)f2}_{1,2,3,4} &=F_{1,3}
 \,=\,F_{1}\,F_{3}-\frac{1}{3}\,(F_{2})^2.
\end{align}
The second equality, between the conformal integral $\mathcal{I}^{(4)f2}$ known as the window diagram and first computed  in \cite{Eden:2016dir},
constitutes the simplest example of the fishnet diagrams computed in \cite{Basso:2017jwq} as determinant of ladders. 

The first equality in eq.~\eqref{eq:NewIdentity} is a new prediction.
The integrals $\mathcal{I}^{(4)d2}$ and $\mathcal{I}^{(4)f}$ are not known analytically for finite cross ratios, however we were able to nontrivially check our prediction thanks to the OPE expansion of these integrals given in the ancillary files of \cite{Chicherin:2018avq}.

A five-loop prediction stemming from a similar argument is recorded in appendix \ref{app:5loopIdentity}.

\subsubsection{Beyond the Basso-Dixon fishnets}

Here we analyze the conformal integrals contributing at leading and subleading loop-order to the partial amplitudes $M_{a,b}$, with $a,b\geq 2$. 

According to \eqref{eq:FishandSubFish}, the leading correction maps to the single determinant in \eqref{eq:leadingS}:
\bba
M_{n+l,n}^{[n(n+l)]}\quad  &= \quad (-1)^{n+l-1}I^{[n(n+l)]}_{n,l}  \nonumber\\
&=\quad (-1)^{(n+1)(l+1)}\,F_{1+{\color{blue}l},\,3+{\color{blue}l},\,\cdots,2n-1+{\color{blue}l}}\,.
\end{align}
On the left-hand side the conditions on the integrand: planarity, exponents $(x_{13}^2)^{n+l}\,(x_{24}^2)^{n}$ and loop order $n(n+l)$, fix it uniquely to be the fishnet diagrams of \cite{Basso:2017jwq}. In order to determine the overall coefficients we analyze the 10-loop data from \cite{Bourjaily:2016evz}. By inspecting the partial amplitudes : 
\beq
\{M_{2,2}^{(4)},\,M_{2,3}^{(6)},\,M_{2,4}^{(8)},\,M_{2,5}^{(10)},\,M_{3,3}^{(9)}\}
\eeq
we conclude that $\left(M_{n+l,n}^{[n(n+l)]} = (-1)^{(n+1)(l+1)}\times\text{fishnet}\right)$ and check the identity:
\bba\label{eq:BDfishnet}
\!\!\!\!\!\!\! \raisebox{-4077787sp}{
  \resizebox{1.7\totalheight}{!}{\includestandalone[width=.3\textwidth]{FigfishnetBassoDixon}}
  }
  &\,=\, F_{1+{\color{blue}l},\,3+{\color{blue}l},\,\cdots,2n-1+{\color{blue}l}}
\,\propto\,\begin{vmatrix}
\color{blue}f_{1+l} & f_{2+l} & \cdots &f_{n+l}  \\
f_{2+l} &  \color{blue}f_{3+l} & \cdots  & f_{n+1+l}\\
\vdots & \vdots & \ddots & \vdots\\
f_{n+l} & f_{n+1+l} & \dots &\color{blue}f_{2n-1+l} \\
\end{vmatrix}
\end{align}
where the proportionality factor is given in \eqref{eq:Mmatrices}. 

This identity between regular fishnets Feynman diagrams and determinants of Ladders was first conjectured in \cite{Basso:2017jwq} and later proven in 
\cite{Derkachov:2019tzo,Derkachov:2020zvv}. For us, this identity appears as a by-product of the relations \eqref{eq:O=M} and \eqref{eq:sumObridge}  between the coulomb branch amplitude and octagons.

\begin{figure}[t]
 \begin{minipage}[h]{1\textwidth}
 \centering 
 \resizebox{4.5\totalheight}{!}{\includestandalone[width=1\textwidth]{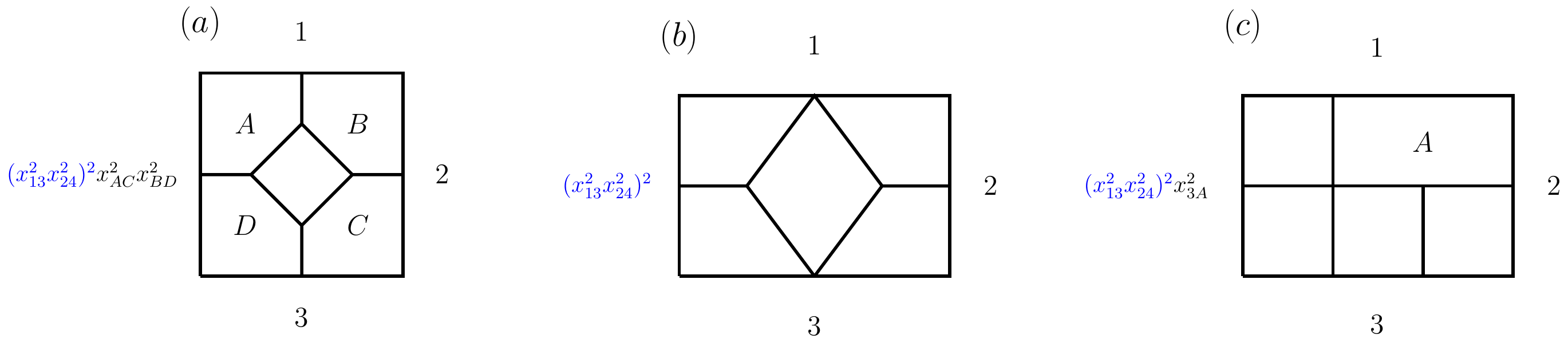}}
 \end{minipage}
 \caption{Five-loop conformal integrals contributing to the next-to-fishnet scattering amplitude $M^{(5)}_{2,2}$}
 \label{fig:fiveloop}
 \end{figure}

The first sub-leading loop order maps to a single determinant according to \eqref{eq:FishandSubFish}:
\begin{align}
M_{n+l,n}^{[n(n+l)+{\color{red}1}]} \,&=\, (-1)^{n+l-1}I^{[n(n+l)+{\color{red}1}]}_{n,l} 
\nonumber\\
&=(-1)^{(n+1)(l+1)}\,2(2n-1+l)\,F_{1+{\color{blue}l},\,3+{\color{blue}l},\,\cdots,\,2n-1+{\color{blue}l}+{\color{red}1}}\,.
\end{align}
The left-hand side is now composed of various conformal integrals. The first instance of these partial amplitudes appears at five loops with $n=2,\,l=0$. It is given by the three integrals represented in fig.~\ref{fig:fiveloop} and its permutations:
 \bba\label{eq:M522}
 M^{(5)}_{2,2}&\,=\,-\,\mathcal{I}^{(5)a}_{1,2,3,4} + \left( \mathcal{I}^{(5)b}_{1,2,3,4} + \mathcal{I}^{(5)b}_{2,1,4,3} \right) - 2\,\left( \mathcal{I}^{(5)c}_{1,2,3,4} + \mathcal{I}^{(5)c}_{2,3,4,1}+ \mathcal{I}^{(5)c}_{3,4,1,2} + \mathcal{I}^{(5)c}_{4,1,2,3} \right)\nonumber\\
   &\,=\,-6\,F_{1,4} \,=
 \,-6\,F_{1}\,F_{4}\,+\, F_{2}\,F_{3}\,.
\end{align}
In order to satisfy the dihedral symmetry the maximum number of permutations is 8, but this is lower when the graph has an extra symmetry. For instance the symmetric graph $\mathcal{I}^{(5)a}$ appears only once, the graph $\mathcal{I}^{(5)b}$ has two inequivalent permutations and the graph $\mathcal{I}^{(5)c}$ has no symmetry so it comes with the maximum number of permutations. After using conformal symmetry these eight graphs reduces to four with an overall factor of 2.
 
 These integrals are unknown for generic values of the cross ratios, but they were computed at leading order in the OPE limit in \cite{Chicherin:2018avq} and we have checked those results are consistent with the prediction on the second line of \eqref{eq:M522}.
 
At higher loops the number of distinct conformal integrals grows as can be seen in the data extracted from \cite{Bourjaily:2016evz} for:
\beq
\{M_{2,3}^{(7)},\,M_{2,4}^{(9)},\,M_{3,3}^{(10)}\}\,.
\eeq
Nevertheless, we were able to identify a simple pattern to construct the correspondent integrands. We conjecture that each integral in $M_{n+l,n}^{[n(n+l)+{\color{red}1}]}$ can be obtained from the $(n+l)\times n$ regular fishnet of \eqref{eq:BDfishnet} by deforming one of its $2\times 2$ sub-lattices in one of the three ways at the top of fig.~\ref{fig:fiveloop} (including the numerators). After generating all integrals, we group them into three families according to which deformation we used and add them up with overall coefficients $\{-1,+1,-1\}$ where the plus sign corresponds to the large diamond deformation $(b)$. This linear combination gives the integrand of $M_{n+l,n}^{[n(n+l)+{\color{red}1}]}$, as it happens in \eqref{eq:M522}.


\section{Discussion}
\label{sec:discussion}

In this paper we showed that a ten-dimensional symmetry unites the correlation functions
of half-BPS scalar operators in four-dimensional $\mathcal{N}$=4 super-Yang-Mills theory
with various R-charge.
This symmetry is realized in a generating function for loop integrands,
which depends exclusively on the 10D sum of spacetime and R-charge space distances
(see eq.~\eqref{eq:Hprediction}).
The symmetry is broken upon integration and
does not simply follow from the 10-dimensional Lagrangian which underlies this theory.
At least up to seven loops, it enables to compute all R-charge correlators from just
the simplest case of stress-tensor correlators, and was confirmed using existing
five-loop results in the literature. 

\begin{figure}[t]
 \begin{minipage}[h]{1\textwidth}
 \centering 
  \resizebox{2\totalheight}{!}{\includestandalone[width=1\textwidth]{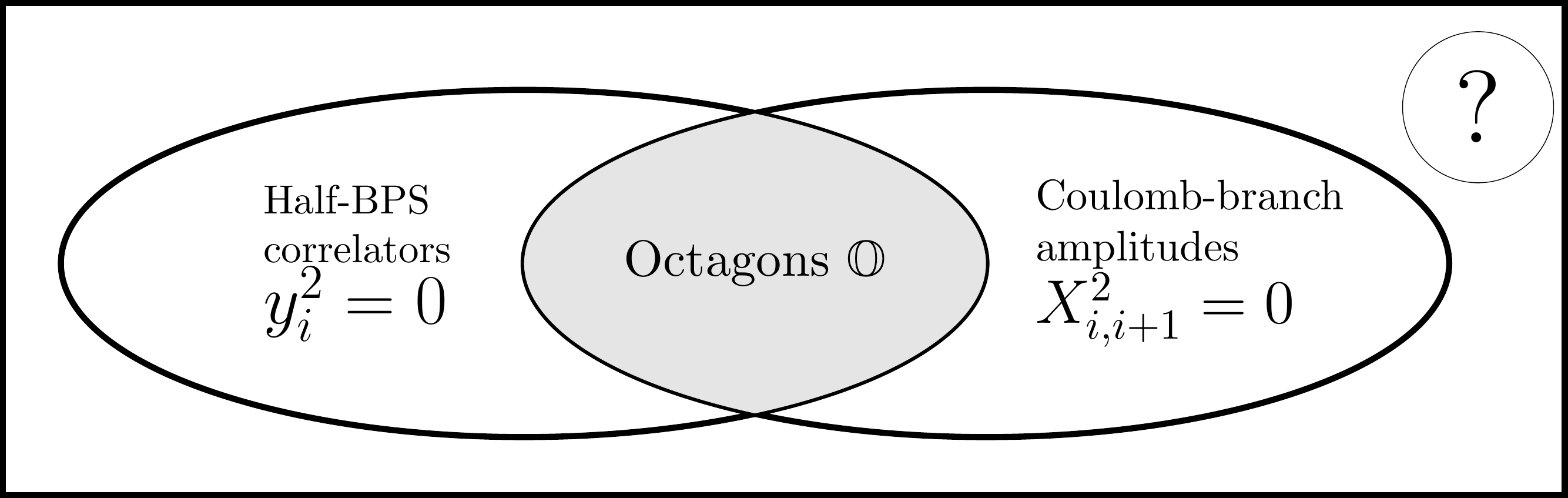}}
 \end{minipage}
 \caption{Half-BPS correlators are labelled by 10D coordinates $(x,y)$ where $y$ is a null 6D vector, while
 Coulomb branch scattering amplitudes are labelled by null-separated 10D points.
 These intersect on the restricted Coulomb branch, on which the octagons are defined.
 We can speculate on the existence of generalized correlators which satisfy a full 10D symmetry
 without any null condition.}
 \label{fig: bigger object}
 \end{figure}

One of the main consequences of this enlarged symmetry is an extension of the correlator/amplitude duality. This was obtained by equating two interpretations of the 10D null limit of the generating function:
as a large R-charge limit giving the so-called octagon of section \ref{sec:FiniteOctagon}, or a scattering amplitude of massive $W$ bosons on the Coulomb branch of the theory.
Since these massive amplitudes are infrared-finite, this equality between correlators and amplitudes holds
not only for the integrand but also at the integrated level.
The duality currently only accesses a strict subset of the kinematic space of each object, which hints at
a possible generalization of the correlation functions which would contain both as depicted in fig.~\ref{fig: bigger object}.
 
Recent integrability-based results make it possible to write the octagon/amplitude as an infinite sum of
well-studied determinants of infinite size (see eq.~\eqref{eq:O10Dnull}).
It is unclear whether the sum could be recast as a single infinite determinant. If possible, this would give,
for the first time, a practical finite-coupling representation for a scattering amplitude in a 4D gauge theory.   

The massless limit of the amplitude on the restricted Coulomb branch is curiously
controlled by an exponent $\Gamma_{\rm oct}$ which differs from the cusp anomalous dimension $\Gamma_{\text{cusp}}$, as discussed in section \ref{4d massless}. Recently these two anomalous dimensions were found to be govern by the same modified BES equation evaluated at two different values of the deformation parameter \cite{Basso:2020xts}.
In light of this development, it is tempting to speculate that this parameter could be associated to different approaches to the massless limit.

Agreement between the octagon integrand and integrability results at weak coupling is nontrivial
and predicts new analytic formulas for certain (sums of) integrals as determinants of ladder integrals (for example, eq.~\eqref{eq:NewIdentity}).  These relations extend the family of regular fishnets to a family of simple linear combinations of deformed fishnets, which also map to single determinants.

Our results are not unrelated to the ten-dimensional
symmetry found in supergravity approximation in ref.~\cite{Caron-Huot:2018kta}.  The objects which satisfy the symmetry are precisely the same: in both cases, the fields $\mathcal{O}(x)$ are packaged into ``pre-ten-dimensional'' ones $\mathcal{O}(x,y)$ by summing all R-charges according to eq.~\eqref{def O} and \eqref{eq:Ok}, and the symmetry is then only satisfied by the
supersymmetrically \emph{reduced} correlator $\mathcal{H}$ (see eq.~\eqref{eq:Hprediction}).

While they apply to the same object, the two ten-dimensional
symmetries are currently not connected: one holds in the weak coupling
limit $g^2_{\rm YM}N_c\to 0$ while the other holds in the supergravity limit
$g^2_{\rm YM}N_c\to \infty$.  In between, the symmetry is not present:
integration breaks the integrand-level symmetry at weak coupling,
while stringy corrections to supergravity are also known to break related spectral degeneracies \cite{Aprile:2020mus}. From the perspective of the present paper, it is hard to understand why the symmetry is restored in the infinite-coupling limit.

What could be the elusive operators in fig.~\ref{fig: bigger object},
which measure $O(x,y)$ when $y^2\neq 0$?
It is tempting to speculate that these could be related to $D$-instantons or $D_{-1}$-branes in the AdS${}_5\times S_5$ geometry. These already appeared in null-separated configurations in \cite{Alday:2007hr,Berkovits:2008ic}. 
To be more precise, the integrands, rather than just a representation of perturbation theory, are best
viewed as exact $(n+\ell)$-point correlators in self-dual Yang-Mills theory \cite{Mason:2010yk,CaronHuot:2010ek}.
We may thus speculate that the 10-dimensional integrands computed in the present paper
are $D_{-1}$ amplitudes in a putative bulk dual of self-dual super Yang-Mills.
Why such correlators enjoy an accidental ten-dimensional conformal symmetry
would still require additional explanation.

\vspace{2cm}
 
\acknowledgments
Work of S.~C.-H. is supported by the National Science and Engineering Council of Canada, the Canada Research Chair program, the Simons Collaboration on the Nonperturbative Bootstrap,
and the Sloan Foundation.
Work of F.~C. is supported in parts by the Canada Research Chair program and the Sloan Foundation.

\appendix 
 
 \section{Definition of 3-loop conformal integrals}
 \label{app:3loops}

 Here we present the basis of 3-loop conformal integrals that appear in the planar four-point functions of  scalar single-trace operators, see \eqref{eq:all3loop}. These are the 3-loop ladder $(L)$, the tennis court $(T)$, the easy  $(E)$ and hard $(H)$ integrals, and finally a product of 1-loop and 2-loop ladder integrals $(g\times h)$ : 
 \bba\label{eq:3loopsCI}
 L_{13;24} &= \frac{(x_{24}^2)^2}{(\pi^2)^3}\int\frac{d^4x_{5}d^{4}x_{6}d^{4}x_{7}}{(x_{15}^2x_{25}^2x_{45}^2)x_{56}^2(x_{26}^2x_{46}^2)x_{67}^2(x_{27}^2 x_{37}^2x_{47}^2)} \, \overset{\text{off-shell}}{=}\, \frac{F_{3}(z,\bar{z})}{x_{13}^2\,x_{24}^2} \,\nonumber\\
T_{13;24} & = \frac{x_{24}^2}{(\pi^2)^3}\int \frac{d^4x_{5}d^4x_{6}d^4x_{7}\,x_{17}^2}{(x_{15}^2x_{25}^2)(x_{16}^2x_{46}^2)(x_{27}^2x_{37}^2x_{47}^2)x_{56}^2x_{57}^2x_{67}^2}\,\overset{\text{off-shell}}{=}\, \frac{F_{3}(z,\bar{z})}{x_{13}^2\,x_{24}^2}\nonumber\\
E_{12;34}&=\frac{x_{23}^2x_{24}^2}{(\pi^2)^3}\int\frac{d^4x_5 d^4x_6 d^4x_7\,x_{16}^2}{(x_{15}^2x_{25}^2 x_{35}^2)x_{56}^2(x_{26}^2x_{36}^2x_{46}^2)x_{67}^2(x_{17}^2x_{27}^2x_{47}^2)}\nonumber\\
H_{12;34} &= \frac{x_{14}^2x_{23}^2x_{34}^2}{(\pi^2)^3}\int\frac{d^4x_5 d^4x_6 d^4x_7\,x_{57}^2}{(x_{15}^2x_{25}^2 x_{35}^2 x_{45}^2)x_{56}^2(x_{36}^2 x_{46}^2)x_{67}^2(x_{17}^2x_{27}^2x_{37}^2x_{47}^2)}\nonumber\\
(g\times h)_{13;24} &= \frac{x_{13}^2 (x_{24}^2)^2}{(\pi^2)^3}\int \frac{d^4x_{5}}{x_{15}^2x_{25}^2x_{35}^2x_{45}^2}\int\frac{d^4x_{6}d^4x_{7}}{(x_{16}^2 x_{26}^2 x_{46}^2)x_{67}^2(x_{27}^2x_{37}^2 x_{47}^2)}\,\overset{\text{off-shell}}{=}\,\frac{F_{1}(z,\bar{z})\,F_{2}(z,\bar{z})}{x_{13}^2 x_{24}^2}
\end{align}
All these integrals are known analytically in terms of single-valued harmonic polylogarithms \cite{Drummond:2013nda}.

\section{A five-loop identity}\label{app:5loopIdentity}

Following the comparison between amplitude integrals and determinants of ladders in section~\ref{sec:comparison}, we find the five-loop identity:
\beq
6\, F_{1,4}\,=\,\raisebox{-14577787sp}{\resizebox{2.3\totalheight}{!}{\includestandalone[width=.6\textwidth]{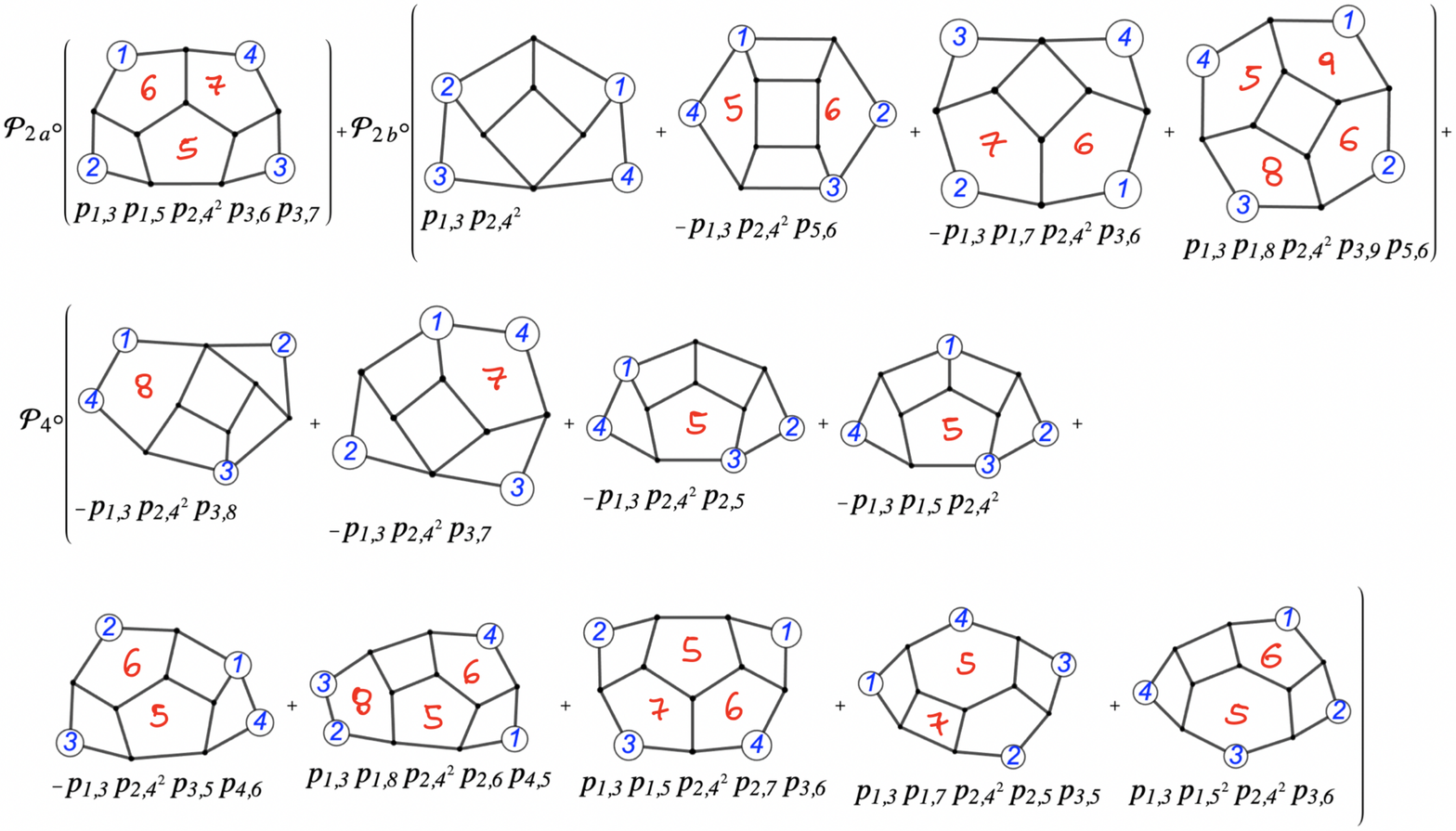}}}\,
\eeq
The numerators of the integrand are recorded below each graph, where  $p_{a,b}\equiv x_{ab}^2$. The correspondent correlation-function graphs are found using graph duality as in  fig.~\ref{fig:3loopAmplitude}.
We use the two-fold and four-fold permutation operators: $\mathcal{P}_{2a}X\equiv X+(1{\leftrightarrow}3)$,
$\mathcal{P}_{2b}X\equiv X+(2{\leftrightarrow}4)$,
and $\mathcal{P}_{4}\equiv \mathcal{P}_{2a}\mathcal{P}_{2b}$.

\newpage
\bibliographystyle{JHEP}
\bibliography{refs}

\end{document}